# Three-dimensional imaging of isolated membrane-protein complexes *in vacuo* with an X-ray laser

Kartik Ayyer[a,b*], Tej Varma Yenupuri[c*], Johan Bielecki[d], Lutz A. Eichacker[e], J. Domingo Meza-Aguilar[f], Louise J. Persson[g], Lena Worbs[c], Patrick Adams[c], Roberto Alvarez[h], Adil Ansari[h], Alfredo Bellisario[i], Rebecca Boll[d], Carl Caleman[i,j], Sebastian Cardoch[i], Andrea Castoldi[k,l], Joe P. J. Chen[m], Cyril Danilevski[d], Merle Dechering[m], Simon Dold[d], Armando D. Estillore[j], Alberto De Fanis[d], Lukas V. Haas[j,b], Patrick Konold[c], David Lomitze[d], Abhishek Mall[a], Dimitra Manatou[h], Iñaki de Diego Martinez[d], Parichita Mazumder[a], Andrew J. Morgan[j], Anna Munke[c], Yevheniy Ovcharenko[d], Matteo Porro[d,n], Safi Rafie-Zinedine[d], Amit K. Samanta[j,b], Christina Schmidt[d], Arezu Sehati[c], Jonas A. Sellberg[o], Björn Senfftleben[d], Carolin Seuring[p,q,r], Zhou Shen[a], Nicusor Timneanu[h], Sergey Usenko[d], Daniel Westphal[c], August Wollter[c], Tamme Wollweber[a,b], Paul Lourdu Xavier[a,d,j], Henry N. Chapman[b,j,s], Janos Hajdu[c,t], Richard A. Kirian[f,h,u], Jochen Küpper[b,j,s], Erik G. Marklund[g], Petra Fromme[f,u], Michael Meyer[d], Filipe R.N.C. Maia[c,v] ✉

[a]Max Planck Institute for the Structure and Dynamics of Matter, Luruper Chaussee 149, 22761 Hamburg, Germany
[b]The Hamburg Centre for Ultrafast Imaging, Universität Hamburg, Luruper Chaussee 149, 22761 Hamburg, Germany
[c]Laboratory of Molecular Biophysics, Department of Cell and Molecular Biology, Uppsala University, Husargatan 3 (Box 596), SE-75124, Uppsala, Sweden
[d]European XFEL, Holzkoppel 4, 22869 Schenefeld, Germany
[e]University of Stavanger, Biochemistry, Postboks 8034, 4068 Stavanger, Norway
[f]Biodesign Center for Applied Structural Discovery, Arizona State University, AZ 85287-5001, Tempe, USA
[g]Department of Chemistry for Life Sciences, Husargatan 3 (Box 576), SE-75123, Uppsala, Sweden
[h]Department of Physics, Arizona State University, Tempe, AZ 85287, USA
[i]Department of Physics and Astronomy, Uppsala University, Box 516, SE-75120, Uppsala, Sweden
[j]Deutsches Elektronen-Synchrotron DESY, Center for Free-Electron Laser Science CFEL, Notkestrasse 85, 22607 Hamburg, Germany
[k]Dipartimento di Elettronica, Informazione e Bioingegneria (DEIB), Politecnico di Milano, Piazza Leonardo da Vinci 32, Milano, 20133, Italy
[l]INFN, Sezione di Milano, Via Celoria 16, Milano, 20133, Italy
[m]Department of Electrical and Computer Engineering, University of Canterbury, Christchurch 8041, New Zealand
[n]Department of Molecular Sciences and Nanosystems, Ca' Foscari University of Venice, 30172 Venezia, Italy
[o]Department of Applied Physics, AlbaNova University Center, KTH Royal Institute of Technology, SE-10691 Stockholm, Sweden
[p]Centre for Structural Systems Biology, Notkestraße 85, 22607, Hamburg, Germany
[q]University of Hamburg, Mittelweg 177, 20148, Hamburg, Germany
[r]Leibniz Institute of Virology, Martinistraße 52, 20251, Hamburg, Germany

[s]Department of Physics, Universität  Hamburg, Luruper Chaussee 149, 22761, Hamburg, Germany
[t]The European Extreme Light Infrastructure, Za Radnici 835, 25241 Dolní Břežany, Czech Republic
[u]School of Molecular Sciences, Arizona State University, AZ 85287-1604, Tempe, USA
[v]NERSC, Lawrence Berkeley National Laboratory, Berkeley, CA, 94720, USA
*These authors contributed equally to this work
✉E-mail: filipe.maia@icm.uu.se

# Abstract

The prospect of imaging single biomolecules, viruses and cells with intense, ultrashort X-ray pulses has driven the development of X-ray free-electron lasers (XFELs). However, the weak scattering from small particles is easily swamped by background from residual gas, which has so far limited applications to strongly scattering targets such as viruses, cell organelles and cells. Here we report a three-dimensional (3D) reconstruction of an isolated 1-MDa membrane-protein complex, photosystem I (PS I), from single-particle diffraction data. PS I trimers were aerosolised by charge-reduction electrospray ionisation and injected into the European XFEL beam, with partial helium gas exchange reducing background scattering by 80%. From 32 788 diffraction patterns of single trimers in random orientations, we reconstructed the 3D electron density to a resolution of 3.8 nm, limited by the detector geometry. The disc-shaped density, about 22 nm across and 10 nm thick, matches the size of a PS I trimer in a detergent micelle and is consistent with the compaction predicted by molecular dynamics simulations of the complex *in vacuo* and observed in native mass spectrometry. These results show that membrane-protein complexes can be imaged *in vacuo* with X-ray lasers, an important step towards ultrafast diffractive imaging of single macromolecules.

# Main

X-ray free-electron lasers (XFELs) are revolutionising structural sciences by providing ultrashort pulses with extremely high intensities. This combination enables probing structures at high spatial and temporal resolution while effectively outrunning radiation damage[1,2]. One application is femtosecond X-ray diffractive imaging (FXI), which has been used to image viruses[3,4], cell organelles[5] and small living cells[6] in the gas phase, and could in principle image individual aerosolised macromolecules[7,8]. Because it requires neither crystals nor embedding in vitreous ice, FXI could make it possible to capture transient conformations and follow ultrafast dynamics once higher resolution is achieved.

Despite its conceptual elegance, FXI presents significant technical challenges when applied to protein-sized samples. The primary difficulty is that such small samples have intrinsically weak scattering cross sections, so their signal is easily overwhelmed by background scattering of the many photons in each XFEL pulse from residual gas and beamline components. Residual gas along the beam path must therefore be minimised to ensure that the recorded signal originates primarily from the molecule of interest. As a result, most FXI experiments to date have focused on larger biological targets, such as cell organelles[5], cells[6], and viruses[3,4], where the scattered signal is stronger and easier to distinguish. While the first diffraction pattern from a single protein was recently recorded[8], the dataset was insufficient to reconstruct a three-dimensional (3D) structure.

Membrane proteins are central to cellular and viral function and are the targets of most drugs[9]. Their dynamic behaviour lets them act as energy converters, transporters, pumps, sensors and signalling hubs. Native mass spectrometry has shown that membrane-protein complexes protected by detergent micelles can be transferred intact from solution into the gas phase, retaining their subunit organisation and bound lipids[10–12]. We studied photosystem I (PS I) from the thermophilic cyanobacterium *Thermosynechococcus vestitus* (formerly *T. elongatus*), a 1-MDa complex of 36 protein subunits and 381 non-covalently bound cofactors. In oxygenic photosynthesis, PS I uses light energy to transfer electrons from plastocyanin or cytochrome $c_6$ to ferredoxin, ultimately generating NADPH. It was also the first protein used to demonstrate serial femtosecond crystallography[13]. Its historical significance and biological importance make PS I a natural model system for FXI studies of individual membrane proteins in the gas phase. Here we report a low-resolution 3D reconstruction of micelle-embedded PS I trimers from gas-phase diffraction data using FXI. The electron density is broadly consistent with the expected outline of a detergent-enveloped trimer, and with higher resolution the same approach could follow the dynamics of individual proteins one particle at a time, unencumbered by crystal packing[14].

PS I trimers were obtained by dissolving PS I crystals[15]. PS I crystallises at low ionic strength, and the crystals were dissolved by raising the ionic strength in the presence of n-dodecyl-β-D-maltoside (DDM), a non-ionic detergent. The resulting protein–detergent micelles protect the complex during the transition from solution to vacuum[10,16]. The particles were aerosolised by charge-reduction electrospray ionisation (ESI)[17] and focused through an aerodynamic lens[18] into the interaction chamber, held at about $10^{-5}$ mbar. There they were intercepted in random orientations by X-ray pulses at the Small Quantum Systems (SQS) instrument of the European XFEL[19], with a photon energy of 1.2 keV (1.03 nm wavelength) and average pulse energy of 5 mJ measured upstream of the beamline optics.

The electrospray produced droplets about 150 nm in diameter, carried in a sheath gas of $N_2$ and $CO_2$ (**Fig. 1**), which were charge-reduced before entering the aerodynamic lens[18]. The solvent evaporated during transport, leaving a thin residue of non-volatile buffer components on the particles (see Methods). Native mass spectrometry nevertheless indicates that structural water molecules remain bound to proteins during electrospray ionisation[20,21]. To reduce background scattering and the K$\alpha$ fluorescence of carbon, nitrogen and oxygen observed in a previous experiment[8], helium was added to the aerosol before the aerodynamic lens[22], replacing most of the $N_2$ and $CO_2$. This reduced the background scattering by 80% (**Fig. 1c**).

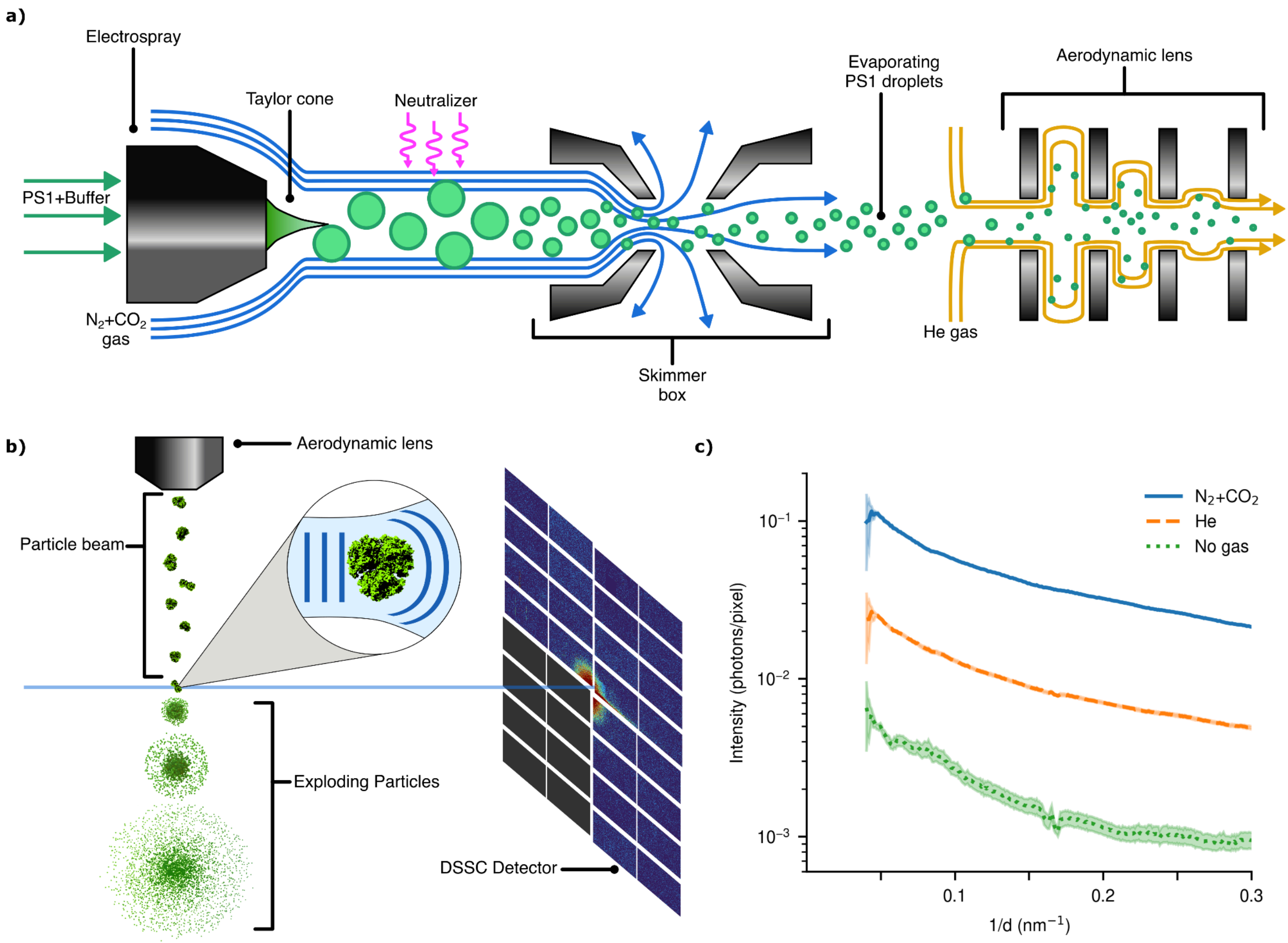

**Fig. 1 | Experimental setup and background scattering. a,** A solution of PS I trimers is aerosolised by electrospray ionisation in a sheath gas of $N_2$ and $CO_2$ at atmospheric pressure. The charged droplets are then charge-reduced by bipolar ions generated from the sheath gas by a vacuum-ultraviolet (VUV) source. Helium is introduced into the low-pressure region of the injector between the skimmer box and the aerodynamic lens, where it replaces most of the $N_2$ and $CO_2$ and so reduces background scattering in the interaction region. **b,** The focused beam of PS I trimers intersects the XFEL pulse train, and the diffraction signal is recorded on a DSSC detector (DSSC, DEPFET sensor with signal compression; DEPFET, depleted p-channel field-effect transistor). **c,** Background scattering for different gas compositions. Replacing most of the $N_2$ and $CO_2$ with helium reduces the background by 80%; the background with no injection gas is shown for reference. Shaded areas show the standard deviation across frames.

Diffraction data were recorded 536 mm downstream of the interaction region on a 1-megapixel DEPFET sensor with signal compression (DSSC) detector[23] (Fig. 1b). The flange holding the detector limits the usable scattering angle to 15.6°, corresponding to a maximum momentum transfer of q = 0.26 $nm^{-1}$ and a resolution of 3.8 nm. One detector quadrant was not operational during the experiment (Fig.

S9). Only a small fraction of X-ray pulses intersect a particle. The remaining frames contain only background scattering from the injection gas and the beamline optics, and were used to estimate the background for each run (see **Methods**).

Of 204 561 630 X-ray pulses, 2 422 532 produced diffraction patterns with signal above background (a hit rate of 1.2%). These patterns were classified into 2D class averages to assess their quality[24]. Classes consistent with a single PS I trimer were seen both edge-on, viewed along the membrane plane (**Fig. 2a**), and face-on (**Fig. 2b**). Other classes showed vertically stacked (**Fig. 2c**) and side-by-side connected (**Fig. 2d**) trimers. Vertical stacking does not occur in the source crystals, so these aggregates cannot result from incomplete dissolution. They most likely form during electrospray and sample delivery, when droplets containing more than one trimer dry and the particles are pushed together, and then remain attached because contact between two surfaces is energetically favoured over exposure to vacuum[10,12].

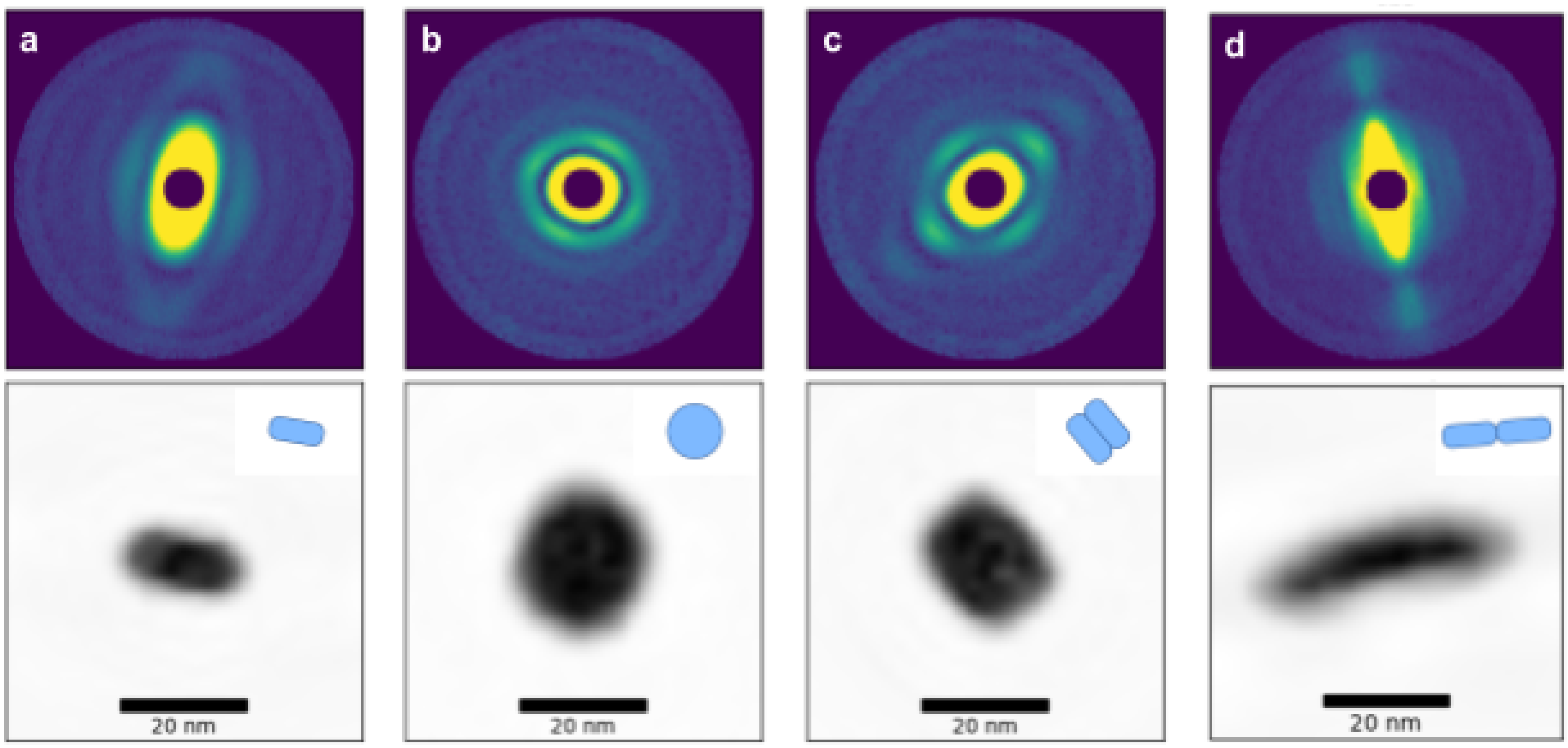

**Fig. 2 | Representative 2D class averages and reconstructed projections.** Top row, 2D class averages of diffraction patterns. Bottom row, projected electron densities obtained from each class average by iterative phase retrieval. **a,** Single trimer, edge-on. **b,** Single trimer, face-on. **c,** Two vertically stacked trimers, edge-on. **d,** Two side-by-side trimers, edge-on. Blue insets show schematically the particle arrangement inferred for each class. Scale bars, 20 nm.

After 2D classification (**Methods** and **Fig. S3**), 32 788 diffraction patterns were attributed to single PS I trimers. These were assembled into a single, self-consistent 3D intensity distribution using the EMC algorithm[25] as implemented in Dragonfly[26], imposing the three-fold rotational symmetry of the PS I trimer. No symmetry was imposed during phasing. The 3D intensities were phased with a combination of the difference map and error reduction algorithms[27,28]. The resulting electron density is disc-shaped, about 22 nm in diameter and 10 nm thick (**Fig. 3**). Projections and slices through the density, interpolated for display, are shown in **Fig. 3a,b**. The phase retrieval transfer function remains above 1/e up to the usable edge of the detector (**Fig. 3c**), giving a resolution of 3.8 nm.

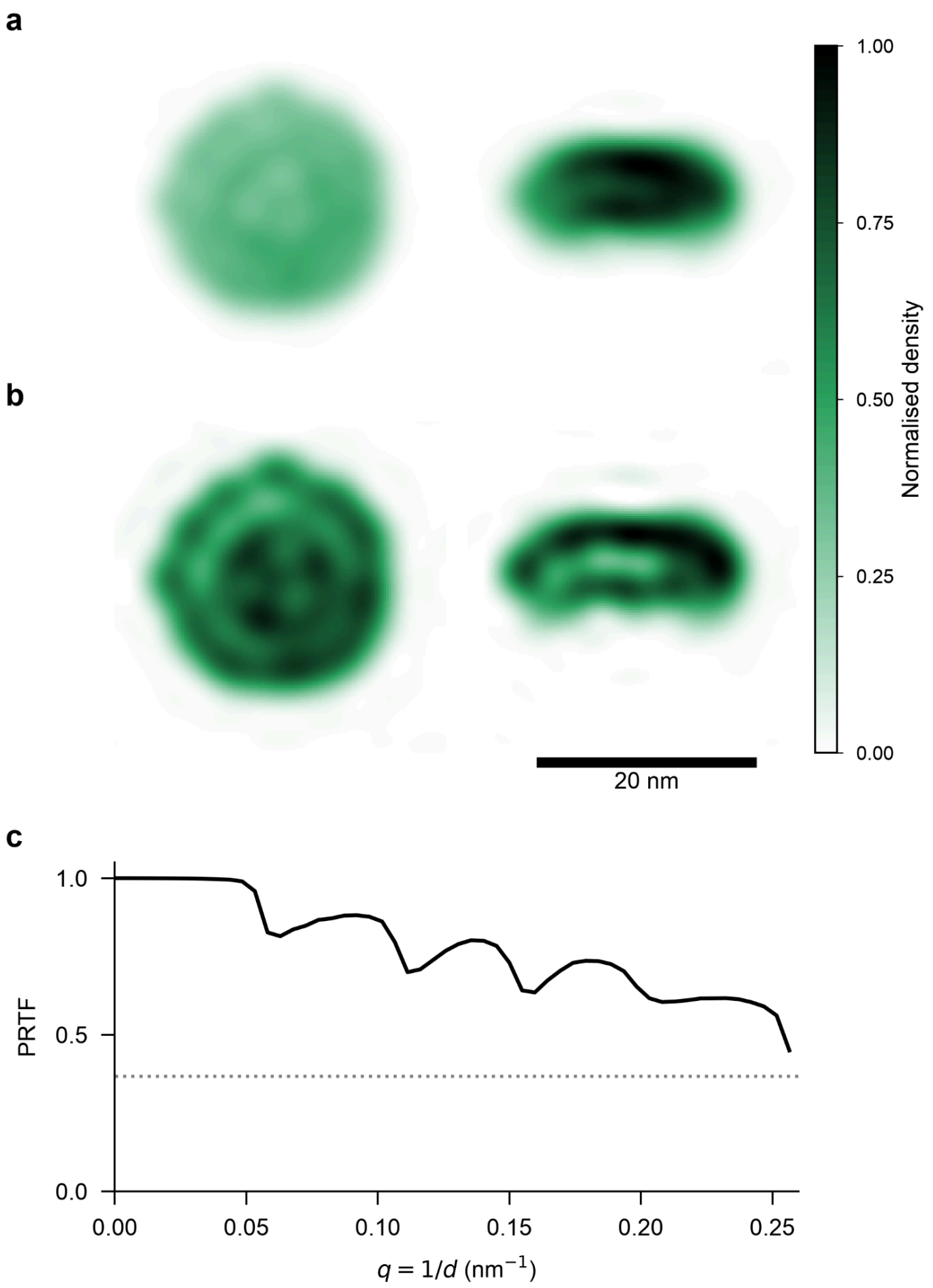

**Fig. 3 | Three-dimensional reconstruction of PS I. a,** Projections of the electron density along the trimer axis (left, face-on) and perpendicular to it (right, edge-on). **b,** Central slices through the electron density in the same orientations. Projected density (**a**) and density (**b**) are each normalised to their maximum value; the map was interpolated for display. Scale bar, 20 nm. **c,** Phase retrieval transfer function (PRTF) from 1000 independent phasing runs. The dotted line marks the 1/e threshold.

Slices through the density reveal three regions of higher density arranged around the centre of the disc (**Fig. 3b**), consistent with the three monomers of the PS I trimer. Three-fold symmetry was imposed on the intensities but not during phasing, so this arrangement is not enforced on the density, although it is favoured by the symmetric intensities. As these features lie close to the resolution limit, they should be interpreted with caution. To assess how the detergent and the vacuum affect the structure, we performed a 1500-ns molecular dynamics simulation of the PS I trimer in a DDM micelle in vacuum. **Fig. 4** compares the experimental map with three models: the crystal structure of the trimer, the trimer in a detergent micelle equilibrated in solution, and the same system after 1500 ns in vacuum. After removal of the solvent, the detergent micelle is rearranged around the protein: over 1500 ns in vacuum, the radius of gyration decreases by about 4% for the whole system and 2.5% for the protein alone (**Fig. 4d,e**), relative to the end of the solution equilibration, reflecting the greater mobility of the detergent compared with the protein. The overall shape and size of the map are visibly closer to this vacuum-equilibrated model than to the solution-equilibrated model (**Fig. 4b,c**). The reconstruction does not, however, show the three protrusions of the stromal ridge that rise above the disc in both the crystal structure and the vacuum-equilibrated model (**Fig. 4**). The spaces between them may be filled by detergent, salt and by water retained on the particle, which may freeze as the droplets cool by evaporation. A similar effect was observed in single-particle diffraction from GroEL, whose central cavity appeared filled with density, although it could not be determined whether this was water, salt or protein[8].

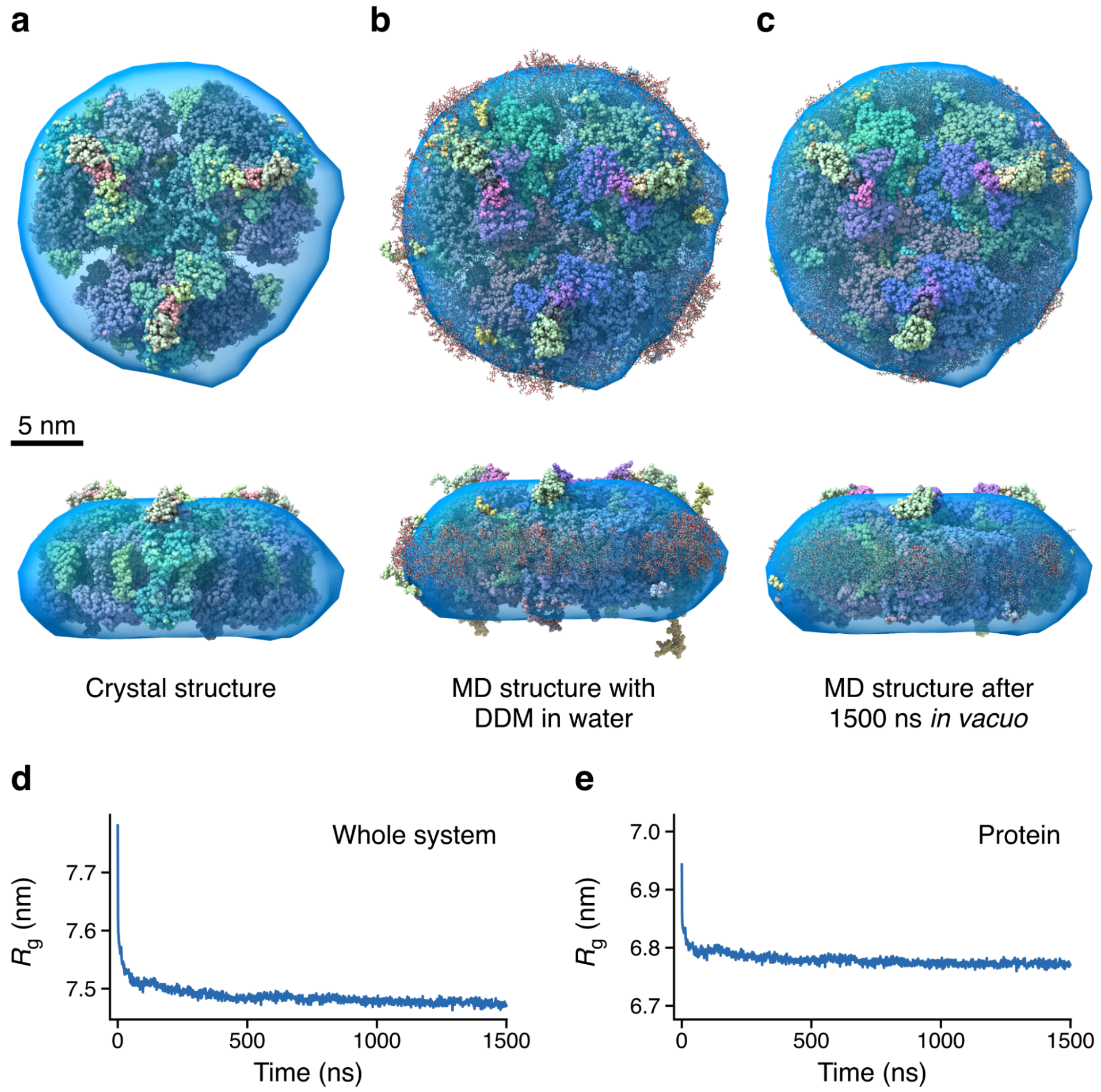

**Fig. 4 | Experimental electron-density map of PS I compared with the crystal structure and two molecular-dynamics models.** The map is shown as a semi-transparent blue isosurface at 0.6 $\varrho_W$. **a,** Crystal structure of PS I including all cofactors (PDB 6PFY[29]). **b,** Molecular dynamics structure of PS I in a DDM detergent micelle after equilibration in

water. **c,** The same PS I–micelle system after 1500 ns of molecular dynamics simulation *in vacuo*. Top row, view along the trimer axis; bottom row, side view with the stromal side up. Scale bar, 5 nm. **d,e,** Radius of gyration (Rg) during the gas-phase simulation for the whole system (PS I, cofactors, detergent and retained water; **d**) and for the protein alone (**e**). Over the simulation, Rg decreases by about 4% for the whole system and 2.5% for the protein, relative to the end of the solution equilibration.

We have reconstructed the 3D structure of a membrane-protein complex from single-particle X-ray diffraction data. The 3.8 nm resolution is modest, but it is set by the usable edge of the detector rather than by the data, as the PRTF remains above threshold out to the highest measured scattering vector. Reaching larger scattering vectors, by raising the photon energy or changing the detector geometry, is therefore the first step towards higher resolution studies. Beyond that, the pulse duration[30], photon energy and focusing optics determine the scattering strength, both directly through the fluence on the sample and indirectly through controlling radiation damage, and their optimal values remain to be established. The effective fluence[31], estimated from the diffraction signal itself without accounting for radiation damage, reached a maximum of about 40 $\mu J/\mu m^2$ in this experiment (see **Fig. S1**). This is lower than the 280 $\mu J/\mu m^2$ measured previously[8], suggesting that the focus, its overlap with the particle beam, or pulse length was not optimal.

Even at this resolution, the reconstruction is consistent with the compaction of the complex *in vacuo*, as predicted by our simulations and observed with native mass spectrometry of membrane proteins[10–12]. Because FXI images particles while they are still in the gas phase, under conditions similar to those of native mass spectrometry and ion mobility experiments, it can complement approaches that image particles after deposition[32,33] and guide the interpretation of results from these fields[10–12,20,21,34]. Further improvements in X-ray optics[35] should bring the resolution to the scale of individual subunits. At that scale, conformational changes triggered by infrared or terahertz pulses, or by pulsed electric fields, could be followed one particle at a time, without the constraints of a crystal lattice or an ice matrix.

# Methods

## Sample purification

PS I trimers were purified from the thermophilic cyanobacterium *Thermosynechococcus vestitus* (formerly *T. elongatus*) following published procedures[29,36], with modifications. Cells were grown at 56 °C in 100-L and 190-L photobioreactors under continuous illumination, supplied with air containing 4% $CO_2$ (50 L/h), yielding approximately 80 g of cells per week. Cells were washed in buffer A (20 mM MES–Na, pH 6.4, 10 mM $MgCl_2$, 10 mM $CaCl_2$), resuspended in buffer A containing 500 mM mannitol, and disrupted in a microfluidiser. Thylakoid membranes were collected by centrifugation and washed four times with buffer A.

Thylakoids at 1 mM chlorophyll were solubilised in the dark for 20 min at 20 °C with 0.6% (w/v) DDM ($\alpha$-isomer content <0.01%), and insoluble material was removed by ultracentrifugation. The extract was loaded onto a DEAE Toyopearl anion-exchange column equilibrated with 5 mM MES, pH 6.4, 20 mM $MgSO_4$ and 0.02% DDM, and eluted with a step gradient of 30%, 50% and 100% high-salt buffer (5 mM MES, pH 6.4, 200 mM $MgSO_4$, 0.02% DDM). PS I trimers eluted in the final, 200 mM step.

The trimer fractions were concentrated to 10 mM chlorophyll in a 400-mL Amicon stirred cell with a 100-kDa cut-off membrane. Crystallisation was then induced by lowering the ionic strength: the concentrate was diluted to 2 mM $MgSO_4$ with salt-free buffer (5 mM MES, pH 6.4, 0.02% DDM) and concentrated again to 10 mM chlorophyll, during which PS I crystals formed. Crystals were recovered in salt-free buffer and stored at 4 °C. Since PS I does not tolerate freezing at any stage of the preparation, the protein was isolated and crystallised shortly before the beamtime.

For the experiments described here, the crystals were sorted by size by timed sedimentation, and the larger crystals were dissolved for injection (see Sample injection). Trimer integrity and sample homogeneity were assessed by transmission electron microscopy and dynamic light scattering (see below).

## Sample injection

Samples were injected with the Uppsala injector[5], which consists of a charge-reduction electrospray ionisation (ESI) unit that aerosolises the sample, followed by an aerodynamic lens that focuses the aerosol into a narrow particle beam and delivers it into the X-ray interaction region under high vacuum.

For injection, PS I crystals were dissolved in 15–20 mM magnesium acetate, pH 7.6, containing 0.01–0.02% DDM, giving PS I trimer concentrations of 0.8–5 mg $mL^{-1}$ ($4.8 \times 10^{14}$ to $3 \times 10^{15}$ particles $mL^{-1}$). Solutions were centrifuged at 20,817 × g for 15 min at 4 °C and kept on ice until use.

The solutions were electrosprayed at atmospheric pressure in a sheath gas of $N_2$ and $CO_2$, at flow rates of 1.0 and 0.15 L $min^{-1}$, respectively[17]. The sample flowed through a fused-silica capillary (30 µm inner diameter, 360 µm outer diameter) whose tip was ground to a cone of approximately 30°. A voltage of 2–3 kV applied to the sample produced a stable Taylor cone[37]. The sample flow rate, set by the pressure in the sample reservoir, was 50–80 nL $min^{-1}$, producing initial droplets of 100–200 nm diameter. The charged droplets were then charge-reduced by bipolar ions generated from the sheath gas by two VUV ionisers (Hamamatsu L12542).

The aerosol entered the injector through two skimmer stages evacuated by scroll pumps, which reduced the pressure from about 1 bar at the electrospray to about 45 mbar after the first skimmer and 0.6–0.7 mbar after the second, at the entrance of the aerodynamic lens. Helium was added between the second skimmer and the lens at about 7.7 mg $min^{-1}$ (≈43 mL $min^{-1}$ at standard conditions), raising the lens entrance pressure to about 2.0 mbar and replacing most of the remaining $N_2$ and $CO_2$; this reduced the background scattering in the interaction region by about 80% (**Fig. 1c**).

The solvent evaporated during transport through the injector, and the aerodynamic lens focused the particles into a narrow beam. Before injecting PS I, the overlap of the particle beam with the X-ray focus was optimised by injecting sucrose particles, using the hit rate as feedback.

**Particle sizing by differential mobility analysis**

Immediately after the XFEL measurements, the size distribution of the aerosolised PS I was measured with a TSI 3080 electrostatic classifier fitted with a TSI 3081 differential mobility analyser, followed by a TSI 3786 condensation particle counter. The aerosol was generated with the same electrospray and charge-reduction settings as in the XFEL experiment and sampled at atmospheric pressure, and particles selected by electrical mobility were counted by the condensation particle counter, covering mobility diameters from 5.83 to 228.8 nm.

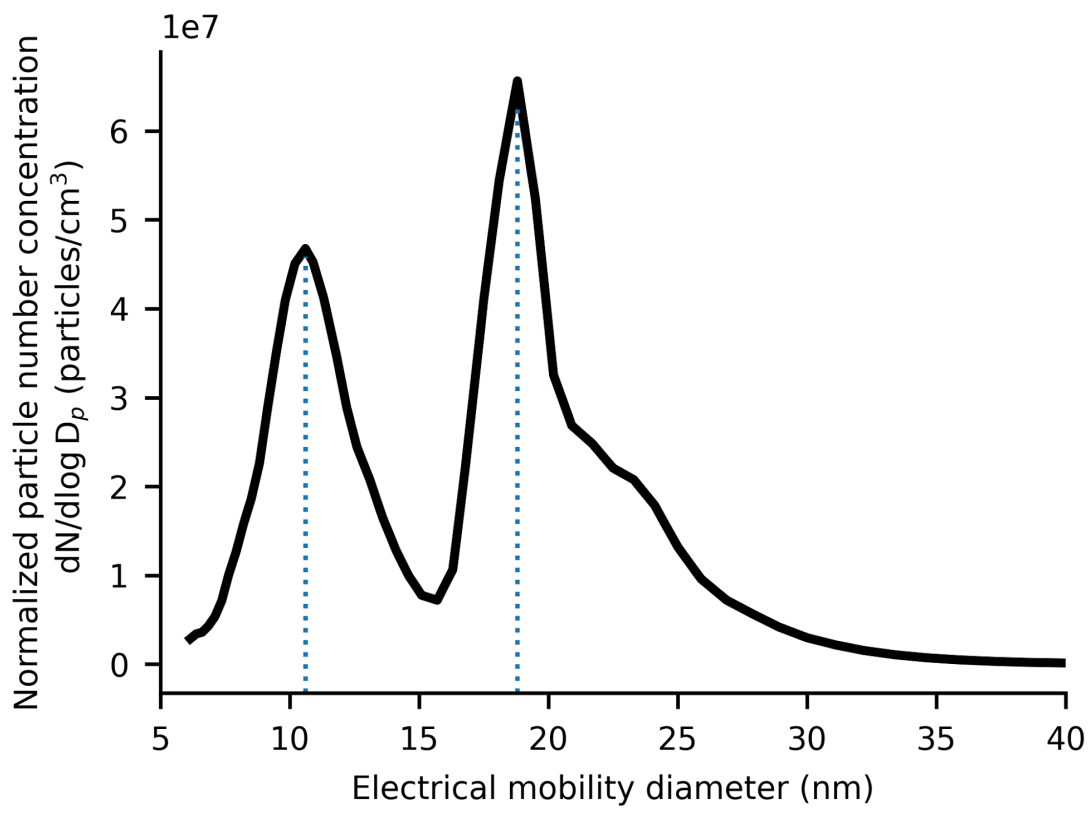

**Fig. 5 | Differential mobility analysis of aerosolised PS I.** Particle number size distribution of the charge-reduced aerosol, generated with the same electrospray and charge-reduction settings as in the XFEL experiment and sampled at atmospheric pressure. Dotted lines mark the two main peaks. The peak at about 18.8 nm is consistent with the electrical mobility diameter calculated from the reconstructed

map (about 20 nm). The peak at about 10.6 nm is attributed to dried droplets that contained non-volatile buffer components but no PS I.

The distribution showed two main populations, centred at about 18.8 nm and 10.6 nm (**Fig. 5**). The electrical mobility diameter calculated for the reconstructed map, using the isosurface shown in Fig. 4, is about 20 nm[38,39], so the 18.8 nm population is consistent with PS I particles. The 10.6 nm population is attributed to dried residue particles of non-volatile buffer components, mainly DDM and magnesium acetate, formed from droplets that contained no PS I.

Assuming that droplets containing PS I carry the same amount of non-volatile material, this material would form a layer about 0.5 nm thick if spread over the surface of a PS I particle. Such a layer lies well below the resolution of the reconstruction and cannot be observed directly, but it is likely to be present.

**Dynamic light scattering**

PS I crystals were dissolved at 1 mg $mL^{-1}$ (about $5.6 \times 10^{14}$ particles $mL^{-1}$) in 15 mM magnesium acetate containing 200 μM DDM (conductivity 1,960 μS $cm^{-1}$), and centrifuged for 10 s at 20,817 × g and 4 °C. Dynamic light scattering was measured at room temperature with a DynaPro NanoStar (Wyatt Technology), and the intensity-weighted distribution of hydrodynamic radii was obtained from ten individual measurements per condition.

Measurements were made on sample remaining after the XFEL experiment, first as prepared and then after adding a further 200 μM DDM (400 μM in total). The additional detergent narrowed the distribution (**Fig. 6**). The dominant population had a hydrodynamic radius of about 10 nm (diameter about 20 nm), consistent with PS I trimers in detergent micelles.

**Transmission electron microscopy of PS I**

PS I at 1 mg $mL^{-1}$ was diluted 100-fold in 15 mM magnesium acetate containing 200 μM DDM. A 3.5 μL aliquot of the resulting ~10 nM PS I solution was applied to 3-4 nm carbon films supported on copper square-mesh grids (CF400-CU, Electron Microscopy Sciences). The grids were glow discharged immediately before sample application using a GloQube Plus system (Quorum Technologies).

After 30 s incubation, the suspension was removed by blotting. The grids were washed twice with water for 1 s, with blotting after each wash. Negative staining was performed using 2% uranyl acetate with two 1 s incubations followed by one 30 s incubation. Excess stain was removed by blotting, and the grids were dried before examination using a JEOL JEM-2100Plus operated at 200 kV and a nominal magnification of 100,000×.

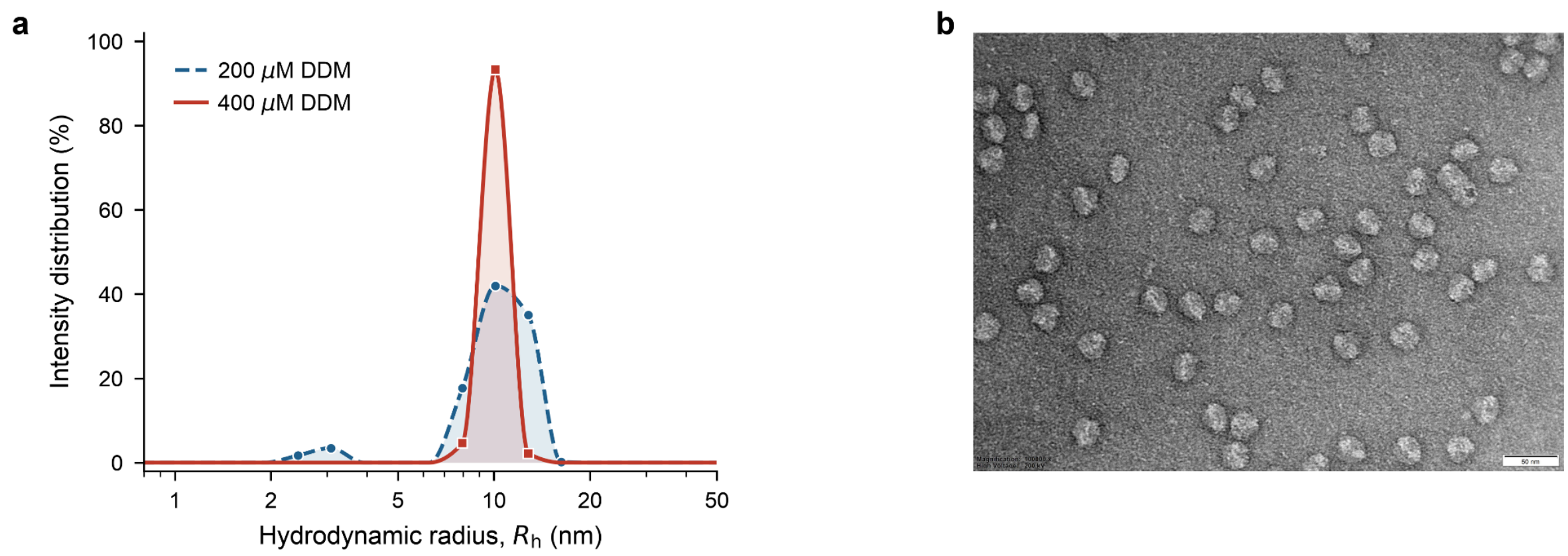

**Fig. 6 | Dynamic light scattering and transmission electron microscopy of PS I. a,** Intensity-weighted distributions of hydrodynamic radius, each from ten measurements, of PS I (1 mg mL⁻¹ in 15 mM magnesium acetate, 200 μM DDM) remaining after the XFEL experiment, as prepared (dashed line) and after adding a further 200 μM DDM (400 μM in total; solid line). **b**, Negative-stain TEM image of PS I from a sample prepared before the XFEL experiment. Scale bar, 50 nm.

## Beamline setup

The European XFEL was tuned to a photon energy of 1.2 keV corresponding to a wavelength of 1.03 nm. The focus of the beam was optimised by maximising the charge states produced when ionising argon, as measured by an ion time-of-flight detector with a 700 mm flight tube. The total energy of each X-ray pulse was measured before any beamline optical element using one of the X-ray gas monitors (XGM) available at the beamline[40], with an average of 5 mJ pulse energy in this study. The accelerator delivered 585 pulses per train at 10 trains per second (2.25 MHz within each train), giving 5,850 pulses per second. The number of pulses per train was limited by the radiation safety limit of 40 W on the maximum beam power in force at the time. In total, 204 million pulses were recorded, and data collection was monitored in real time with *Hummingbird*[41].

## Detector processing and hit finding

The DSSC detector can record up to 800 frames per pixel per train[23]. For the data described here, 585 frames were collected per train. Each memory cell was calibrated individually to convert the digitised signal into photon counts, accounting for gain variations such as slow drifts and even/odd differences arising from mismatched pipelined sample-and-hold capacitors. Pedestals were determined first: dark runs were recorded at intervals without X-rays, and the mean and standard deviation were calculated for each memory cell.

Next, the gain, i.e. the signal per photon, was estimated. The average gain was set to about 5 analogue-to-digital units (ADU) per 1.2 keV photon. Gain variations were determined from histograms of offset-corrected signals in 88,000 frames containing X-ray diffraction. The gain varied mainly between pixel columns within each 64×64-pixel application-specific integrated circuit (ASIC), and was stable within a column across memory cells and between ASICs. A gain was therefore assigned to each column (Fig. 7).

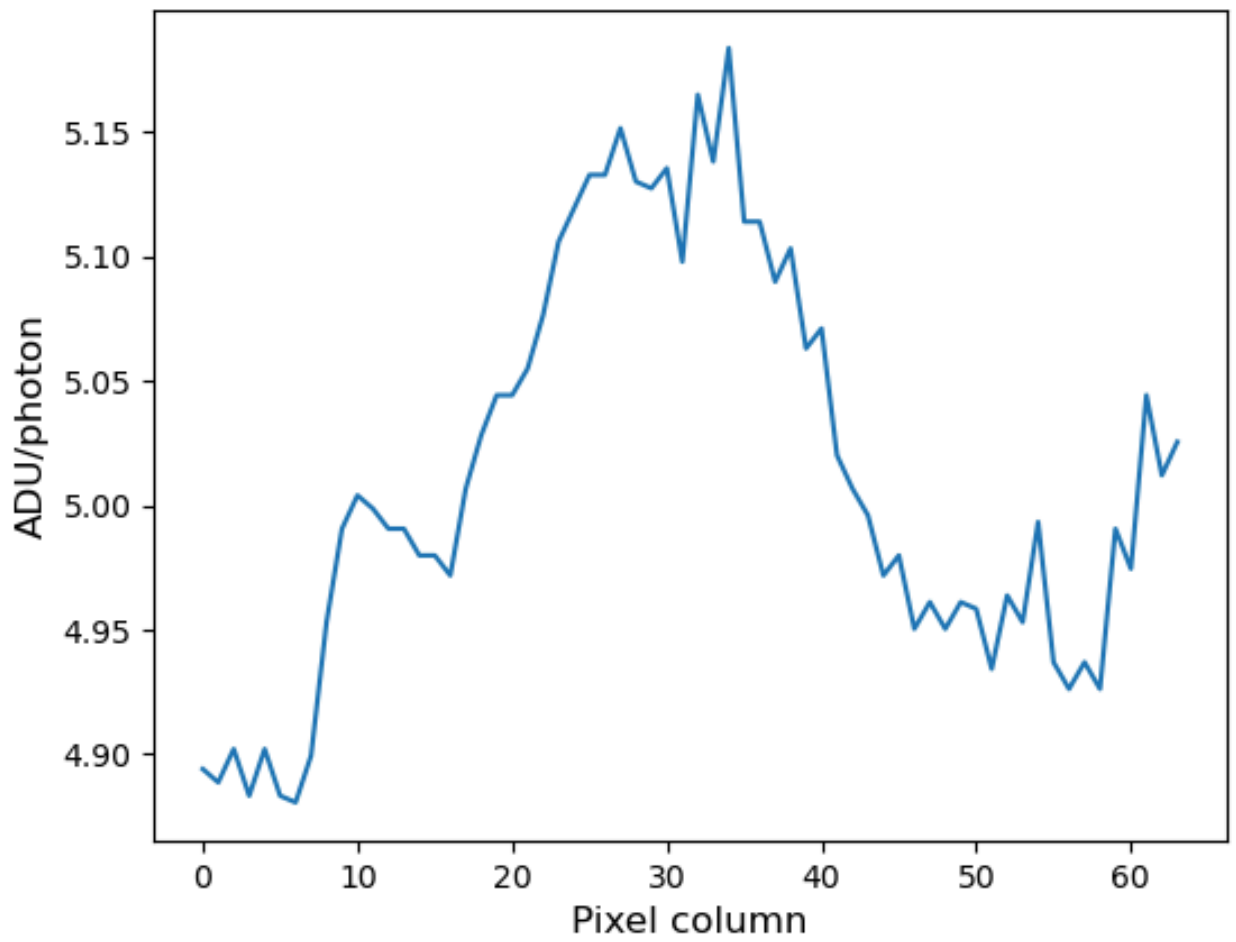

**Fig. 7 | Calibration of the detector.** The conversion factor between ADUs and photons as a function of the pixel column.

The offset- and gain-corrected signals were converted to photon counts using a threshold that depends on the standard deviation $\sigma$ of each memory cell. Following the method described in Ayyer et al. 2021[24], for an expected signal of $10^{-3}$ photons per pixel per frame, the threshold was

$$t = 0.5 - (\sigma/5)^2 \ln(10^{-3})$$

where $5$ corresponds to the nominal gain in ADU per photon. Pixels with values below $t$ were assigned zero photons, and values between $t + k - 1$ and $t + k$ were assigned $k$ photons. Pixels in the shadow of the detector flange, in the non-operational quadrant and around the direct beam were masked (Fig. S9).

Hits were identified as outliers in the number of lit pixels per frame. Because this was done on all frames before full photon conversion, a lit pixel was defined as one whose dark-corrected signal exceeded 3.75 ADU, three-quarters of the nominal signal per photon. For each 5-minute run (1.75 million frames), a Gaussian was fitted to the background peak of the lit-pixel distribution, and frames whose lit-pixel count exceeded the mean of this background by more than four standard deviations were classified as hits (**Fig. 8**).

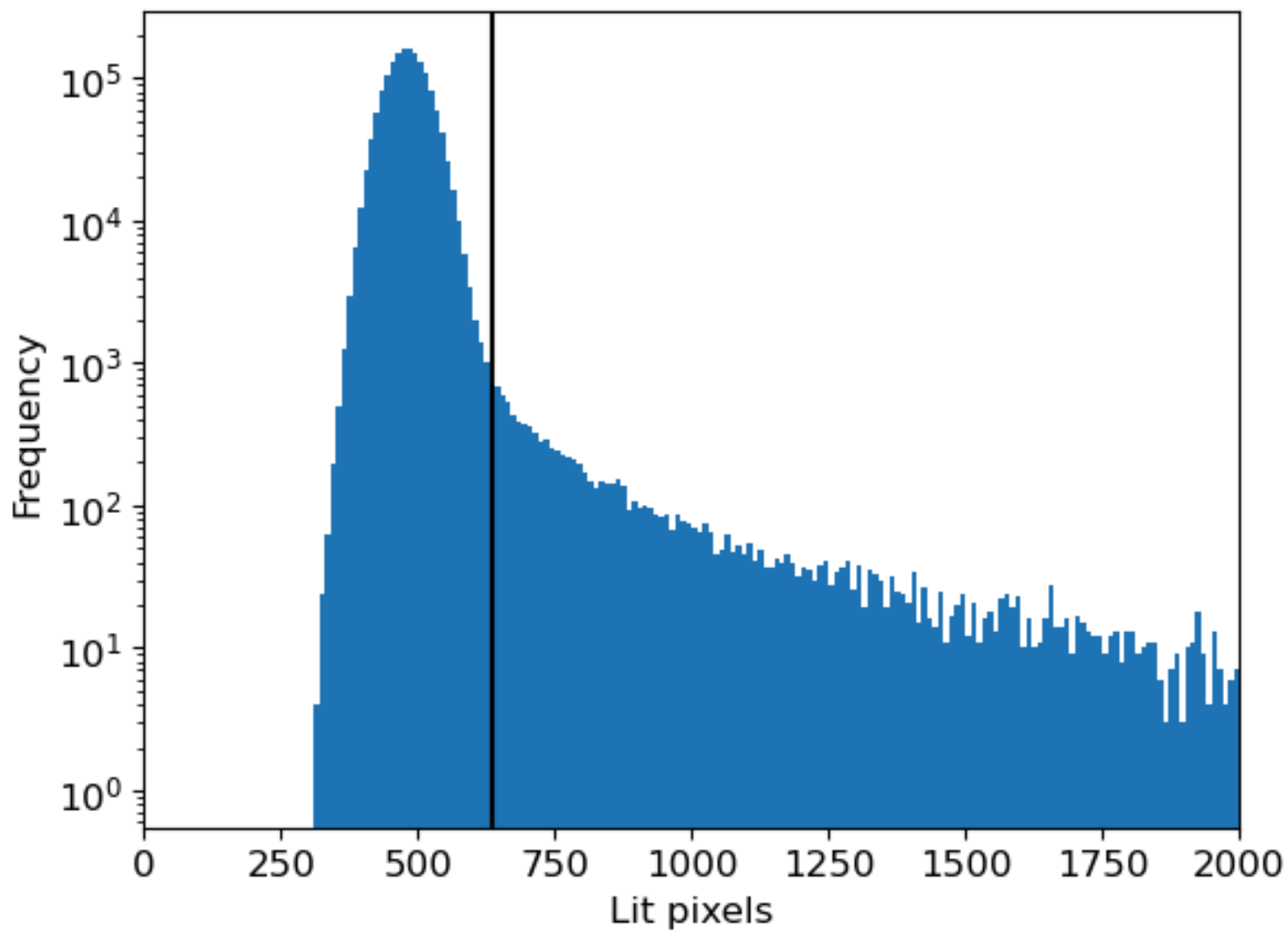

**Fig. 8 | Hit finding.** Distribution of the number of lit pixels per frame (pixels with a dark-corrected signal above 3.75 ADU) for a typical 5-minute run of about 1.75 million frames, on a logarithmic scale. The peak contains background frames; the tail to the right contains frames with diffraction from particles. The vertical line marks the hit threshold, four standard deviations above the mean of a Gaussian fitted to the background peak. Frames to its right were classified as hits.

## 2D classification and 3D intensity reconstruction

Patterns from single PS I trimers were selected from the 2 422 532 hits by successive rounds of 2D classification in *Dragonfly*, each with 50 classes. The selection is summarised in **Fig. S3**, and the class averages from each step are shown in **Figs S4–S8**. Successive rounds removed weak hits and detector artefacts (**Fig. S4**), obvious multiple hits (**Fig. S5**), stacked trimers, and other particles larger than a single micelle-embedded trimer (**Fig. S6**).

The remaining patterns were then split by orientation. Top views, with the disc normal to the beam, give nearly isotropic patterns in which a single trimer differs from larger particles mainly in fringe spacing. These were classified with six-fold symmetry imposed, as expected from the trimer's three-fold axis combined with the centrosymmetry of the diffraction intensities, and only the classes with the smallest particles were kept (**Fig. S7**). All other orientations underwent a final round of background-aware classification, which removed weak hits and thick discs, since at this resolution a tilted single trimer and an edge-on stacked pair give similar patterns (**Fig. S8**). Together, the two selections gave the 32 788 patterns used for 3D reconstruction (**Fig. S2**).

Background-aware classification and 3D intensity reconstruction used a modified version of the EMC algorithm[42], implemented in *Dragonfly*, which incorporates a known background for each frame but treats the incident fluence on each particle as unknown. Backgrounds were estimated from the frames without hits ("misses") in each 5-minute run. The 3D intensities were assembled with three-fold rotational symmetry imposed.

Phases were retrieved with a combination of the error reduction and difference map algorithms, using a fixed support. No symmetry was imposed during phasing. Before phasing, a 3D background was subtracted using a rolling minimum filter 1.5 times the fringe spacing in width, to avoid spurious bright spots in the reconstructed density[43]. One thousand independent reconstructions were averaged, after alignment including with the centrosymmetric inverses. The resulting phase retrieval transfer function is shown in **Fig. 3c**.

For display and measurement, the averaged map was interpolated 9-fold, by Fourier interpolation. Isosurfaces are shown at 0.6 $\varrho_w$, where $\varrho_w = \langle\varrho^2\rangle/\langle\varrho\rangle$ is the density-weighted mean density and $\varrho$ is the electron density. Dimensions were measured along the principal axes of the largest connected region above this threshold.

## Preparation of simulation model

The PS I simulation model was based on an X-ray crystal structure of the complex (PDB ID: 6PFY[29]), with missing residues reconstructed using MODELLER[44]. The protein was embedded into a micelle consisting of 500 DDM molecules using the CHARMM-GUI Micelle Builder[45–47] and PPM 2.0[48]. Following embedding, cofactors were reintroduced to the system. Incomplete cofactor structures were refined using PyMOL's building and sculpting tools, and some chlorophyll molecules were removed to resolve steric clashes with the detergent molecules. The final assembled system comprised 258 chlorophyll, 66 β-carotene, 9 1,2-dipalmitoyl-phosphatidyl-glycerole (LHG), 3 1,2-distearoyl-monogalactosyl-diglyceride (LMG), 9 $[Fe_4S_4]^{2+}$ clusters and 6 phylloquinone molecules.

## Force field parameters

Simulations were performed using the CHARMM36 all-atom force field[49]. Parameters for the protein and DDM were utilised from the standard CHARMM36 distribution. Parameters for β-carotene and phylloquinone were generated via the CGenFF web tool, while parameters for the lipids, LMG and LHG, were obtained through the CHARMM-GUI Ligand Reader & Modeler[50]. Chlorophyll parameters were taken from Guerra et al. 2015[51], and the cysteine-coordinated $[Fe_4S_4]^{2+}$ clusters were taken from da Silva et al. 2022[52]. Virtual sites not present in the base force field were generated using MkVsites to allow for a longer timestep[53].

## Simulation protocol

Simulations of the DDM-embedded PS I complex were performed using an implementation of GROMACS based on version 2019[54] that incorporates the fast multipole method (FMM)[55]. First, the system was equilibrated in explicit solvent using the TIP3P water model[56]. These simulations placed the system in a rhombic dodecahedral box with a minimum distance of 1 nm between the sides of the box and the solute, and employed periodic boundary conditions in all directions. The equilibration followed a multi-step protocol, outlined in Table 1.

Energy minimisation runs with steepest descent ran until machine precision was reached. To prevent instabilities related to the $[Fe_4S_4]^{2+}$ clusters, their coordinates were frozen during energy minimisation, and the bond force constants connecting the Fe ions to cysteine sulfur atoms were gradually increased to their target values. Temperature-coupled simulations used the v-rescale thermostat[57], coupling the solute (protein and cofactors), detergent, and solvent (water and ions) separately at 300 K with a coupling constant of 1.0 ps. They also employed translational center-of-mass motion removal, applied to the solute and detergent as one group, and to the solvent separately. LINCS constraints of the 6th order were applied to all bonds, using one iteration. Coulomb forces were computed with particle mesh Ewald[58], and van der Waals forces were computed using a 1.2-nm cutoff and the Force-switch modifier with a 1.0-nm cutoff. The Verlet scheme was employed for neighbour searching, updated every 20th step. All runs applying pressure coupling used isotropic coupling with the C-rescale barostat set to 1.0 bar and a compressibility of 4.5e-5 $bar^{-1}$, with a coupling constant of 5.0 ps. Reference coordinates for position restraints were scaled based on the center of mass, and translational center-of-mass motion was removed at 100-step intervals, coupling the protein, detergent, and cofactors separately from the solvent.

Simulation of the PS I complex in solution caused a small gap to form in the DDM micelle. To prevent this from impacting the subsequent vacuum simulations, additional DDM molecules were added to fill the gap, the system was re-solvated, and then underwent a final solution equilibration, which is also outlined in Table 1. The final system used in subsequent vacuum simulations contained 508 DDM molecules.

**Table 1 | Parameters for simulations in solution.** Restraints of positions (*) and dihedrals (**) were applied to heavy atoms only.

| | Energy minimisation | Equilibration | Production | Final equilibration |
|---|---|---|---|---|

| Integrator | Steep | | | | MD | | | | | | MD | Steep | MD | MD |
|---|---|---|---|---|---|---|---|---|---|---|---|---|---|---|
| Protein* | 1,000 | 750 | 500 | 250 | 2,000 | 1,000 | 500 | 250 | 50 | 0 | 0 | 250 | 250 | 250 |
| DDM* | 1,000 | 1,000 | 1,000 | 1,000 | 1,000 | 0 | 0 | 0 | 0 | 0 | 0 | 1,000 | 1,000 | 0 |
| DDM** | 1,000 | 1,000 | 1,000 | 1,000 | 1,000 | 400 | 200 | 200 | 100 | 0 | 0 | 1,000 | 1,000 | 200 |
| $Fe_4S_4$ frozen | Yes | Yes | Yes | Yes | No | No | No | No | No | No | No | Yes | No | No |
| Fe-Cys kb ( | 100 | 100 | 100 | 1000 | Full strength | Full strength | Full strength | Full strength | Full strength | Full strength | Full strength | Full strength | Full strength | Full strength |
| Number of steps | - | - | - | - | 125,000 | 125,000 | 125,000 | 250,000 | 250,000 | 250,000 | 500,000 | - | 125,000 | 125,000 |
| Time step (ps) | - | - | - | - | 0.001 | 0.001 | 0.001 | 0.002 | 0.002 | 0.002 | 0.005 | - | 0.001 | 0.001 |
| Pressure coupling | - | - | - | - | No | No | Yes | Yes | Yes | Yes | Yes | - | No | Yes |

To prepare the complex for gas-phase simulation, it was neutralised by adjusting the protonation state of negatively charged residues. Residues with the highest solvent-exposed surface area were selected for neutralisation, determined by calculating residue-wise solvent-accessible surface area on the solvent-equilibrated structure using MDTraj[59]. A total of 274 water molecules were retained at positions corresponding to the crystallographic waters identified in the original X-ray structure. The system was placed in a cubic simulation box with a minimum distance of 3 nm between the system and the sides of the box.

The gas-phase simulation protocol is outlined in Table 2. These simulations utilised FMM for the calculation of Coulomb forces[55], using settings for the tree depth (*d*) and multipole order (*p*) as established previously[60]. Van der Waals interactions were treated with a 1.0-nm cutoff, and the potential-shift modifier. Both translational and rotational center-of-mass motion for the whole system were removed every 10 steps. Simulations applying temperature control used the v-rescale thermostat and a coupling constant of 0.1 ps, with the detergent coupled separately from the rest of the system. All bonds were constrained using 6th-order LINCS, with one iteration. Due to the vacuum-like conditions, no periodic boundary conditions were used, and no barostat.

**Table 2. Parameters for simulations in the gas phase.** Restraints of positions (*) and dihedrals (**) were applied to heavy atoms only.

| | Energy minimisation | Equilibration | | | Production |
|---|---|---|---|---|---|
| Integrator | Steep | MD | | | MD |
| Protein* | 1,000 | 1,000 | 1,000 | 0 | 0 |
| DDM* | 1,000 | 1,000 | 1,000 | 0 | 0 |
| DDM** | 1,000 | 1,000 | 1,000 | 0 | 0 |
| Water* | 1,000 | 1,000 | 1,000 | 0 | 0 |
| Temperature (K) | - | 150 | 300 | 300 | 300 |
| Number of steps | - | 250,000 | 250,000 | 250,000 | 300,000,000 |
| Time step (ps) | - | 0.002 | 0.002 | 0.002 | 0.005 |
| FMM,d | 0 | 4 | 4 | 4 | 4 |
| FMM,p | 0 | 8 | 8 | 8 | 8 |

## Data availability

A total of 4.8 million detector images were deposited on the Coherent X-ray Imaging Data Bank (CXIDB)[61] under ID 247, together with the reconstructed map. This includes sample runs (2.4 million images), detector calibration runs (1.1 million images), runs with only the X-ray beam (200 thousand images) and X-ray beam, sample delivery gas but without sample (1.1 million images). The DOI for the original data at the European XFEL is https://doi.org/10.22003/XFEL.EU-DATA-004098-00. Files related to the molecular dynamics simulation have been deposited at https://doi.org/10.5281/zenodo.23079190.

## Code availability

The data analysis scripts used are available from https://github.com/FXIhub/xfel4098.

# References

1. Neutze, R., Wouts, R., van der Spoel, D., Weckert, E. & Hajdu, J. Potential for biomolecular imaging with femtosecond X-ray pulses. *Nature* **406**, 752–757 (2000).
2. Chapman, H. *et al.* Femtosecond diffractive imaging with a soft-X-ray free-electron laser. *Nat. Phys.* **2**, 839–843 (2006).
3. Seibert, M. M. *et al.* Single mimivirus particles intercepted and imaged with an X-ray laser. *Nature* **470**, 78–81 (2011).
4. Ekeberg, T. *et al.* Three-dimensional reconstruction of the giant mimivirus particle with an x-ray free-electron laser. *Phys. Rev. Lett.* **114**, 098102 (2015).
5. Hantke, M. F. *et al.* High-throughput imaging of heterogeneous cell organelles with an X-ray laser. *Nat. Photonics* **8**, 943–949 (2014).

6. van der Schot, G. *et al.* Imaging single cells in a beam of live cyanobacteria with an X-ray laser. *Nat. Commun.* **6**, 5704 (2015).
7. Hajdu, J. *et al. LCLS: The First Experiments.* 35–62 https://ntrl.ntis.gov/NTRL/dashboard/searchResults/titleDetail/DE2003812649.xhtml (2000).
8. Ekeberg, T. *et al.* Observation of a single protein by ultrafast X-ray diffraction. *Light Sci. Appl.* **13**, 15 (2024).
9. Overington, J. P., Al-Lazikani, B. & Hopkins, A. L. How many drug targets are there? *Nat. Rev. Drug Discov.* **5**, 993–996 (2006).
10. Barrera, N. P., Di Bartolo, N., Booth, P. J. & Robinson, C. V. Micelles Protect Membrane Complexes from Solution to Vacuum. *Science* **321**, 243–246 (2008).
11. Laganowsky, A. *et al.* Membrane proteins bind lipids selectively to modulate their structure and function. *Nature* **510**, 172–175 (2014).
12. Gupta, K. *et al.* The role of interfacial lipids in stabilizing membrane protein oligomers. *Nature* **541**, 421–424 (2017).
13. Chapman, H. N. *et al.* Femtosecond X-ray protein nanocrystallography. *Nature* **470**, 73–77 (2011).
14. Konold, P. E. *et al.* Confinement in crystal lattice alters entire photocycle pathway of the Photoactive Yellow Protein. *Nat. Commun.* **11**, 4248 (2020).
15. Jordan, P. *et al.* Three-dimensional structure of cyanobacterial photosystem I at 2.5 Å resolution. *Nature* **411**, 909–917 (2001).

16. Kumar, S. *et al.* Native Mass Spectrometry of Membrane Protein–Lipid Interactions in Different Detergent Environments. *Anal. Chem.* **96**, 16768–16776 (2024).

17. Bielecki, J. *et al.* Electrospray sample injection for single-particle imaging with x-ray lasers. *Sci. Adv.* **5**, eaav8801 (2019).

18. Hantke, M. F. *et al.* Rayleigh-scattering microscopy for tracking and sizing nanoparticles in focused aerosol beams. *IUCrJ* **5**, 673–680 (2018).

19. Tschentscher, T. *et al.* Photon beam transport and scientific instruments at the european XFEL. *Appl. Sci.* **7**, 592 (2017).

20. Sobott, F., Hernández, H., McCammon, M. G., Tito, M. A. & Robinson, C. V. A Tandem Mass Spectrometer for Improved Transmission and Analysis of Large Macromolecular Assemblies. *Anal. Chem.* **74**, 1402–1407 (2002).

21. Ruotolo, B. T. *et al.* Evidence for Macromolecular Protein Rings in the Absence of Bulk Water. *Science* **310**, 1658–1661 (2005).

22. Yenupuri, T. V. *et al.* Helium-electrospray improves sample delivery in X-ray single-particle imaging experiments. *Sci. Rep.* **14**, 4401 (2024).

23. Porro, M. *et al.* The MiniSDD-based 1-Megapixel Camera of the DSSC Project for the European XFEL. *IEEE Trans. Nucl. Sci.* 1–1 (2021) doi:10.1109/TNS.2021.3076602.

24. Ayyer, K. *et al.* 3D diffractive imaging of nanoparticle ensembles using an x-ray laser. *Optica* **8**, 15 (2021).

25. Loh, N.-T. D. & Elser, V. Reconstruction algorithm for single-particle diffraction imaging experiments. *Phys. Rev. E Stat. Nonlin. Soft Matter Phys.* **80**, 026705 (2009).

26. Ayyer, K., Lan, T.-Y., Elser, V. & Loh, N. D. Dragonfly: an implementation of the expand-maximize-compress algorithm for single-particle imaging. *J. Appl. Crystallogr.* **49**, 1320–1335 (2016).

27. Elser, V. Solution of the crystallographic phase problem by iterated projections. *… Sect. Found. Crystallogr.* 201–209 (2003).

28. Ayyer, K. *et al.* Low-signal limit of X-ray single particle diffractive imaging. *Opt. Express* **27**, 37816–37833 (2019).

29. Gisriel, C. *et al.* Membrane protein megahertz crystallography at the European XFEL. *Nat. Commun.* **10**, 5021 (2019).

30. Khubbutdinov, R. *et al.* High spatial coherence and short pulse duration revealed by the Hanbury Brown and Twiss interferometry at the European XFEL. *Struct. Dyn.* **8**, 044305 (2021).

31. You, T., Bielecki, J. & Maia, F. R. N. C. Impact of gas background on XFEL single-particle imaging. *Sci. Rep.* **15**, 29559 (2025).

32. Esser, T. K. *et al.* Cryo-EM of soft-landed β-galactosidase: Gas-phase and native structures are remarkably similar. *Sci. Adv.* **10**, eadl4628 (2024).

33. Eriksson, L. *et al.* High-resolution cryoEM structure determination of soluble proteins after soft-landing electrospray ion beam deposition. Preprint at https://doi.org/10.48550/arXiv.2503.22364 (2026).

34. Bolla, J. R., Agasid, M. T., Mehmood, S. & Robinson, C. V. Membrane Protein–Lipid Interactions Probed Using Mass Spectrometry. *Annu. Rev. Biochem.* **88**, 85–111 (2019).

35. Yamada, J. *et al.* Extreme focusing of hard X-ray free-electron laser pulses enables 7 nm focus width and 1022 W cm−2 intensity. *Nat. Photonics* **18**, 685–690 (2024).

36. Fromme, P. & Witt, H. T. Improved isolation and crystallization of photosystem I for structural analysis. *Biochim. Biophys. Acta BBA - Bioenerg.* **1365**, 175–184 (1998).

37. Gañán-Calvo, A. M. & Montanero, J. M. Revision of capillary cone-jet physics: electrospray and flow focusing. *Phys. Rev. E Stat. Nonlin. Soft Matter Phys.* **79**, 066305 (2009).

38. Kim, J. H., Mulholland, G. W., Kukuck, S. R. & Pui, D. Y. H. Slip Correction Measurements of Certified PSL Nanoparticles Using a Nanometer Differential Mobility Analyzer (Nano-DMA) for Knudsen Number From 0.5 to 83. *J. Res. Natl. Inst. Stand. Technol.* **110**, 31–54 (2005).

39. Mackowski, D. W. Monte Carlo simulation of hydrodynamic drag and thermophoresis of fractal aggregates of spheres in the free-molecule flow regime. *J. Aerosol Sci.* **37**, 242–259 (2006).

40. Maltezopoulos, T. *et al.* Operation of X-ray gas monitors at the European XFEL. *J. Synchrotron Radiat.* **26**, 1045–1051 (2019).

41. Daurer, B. J., Hantke, M. F., Nettelblad, C. & Maia, F. R. N. C. Hummingbird: monitoring and analyzing flash X-ray imaging experiments in real time. *J. Appl. Crystallogr.* **49**, 1042–1047 (2016).

42. Lan, T.-Y. *et al.* Solving protein structure from sparse serial microcrystal diffraction data at a storage-ring synchrotron source. *IUCrJ* **5**, 548–558 (2018).

43. Lundholm, I. V. *et al.* Considerations for three-dimensional image reconstruction from experimental data in coherent diffractive imaging. *IUCrJ* **5**, 531–541 (2018).

44. Sali, A. & Blundell, T. L. Comparative protein modelling by satisfaction of spatial restraints. *J. Mol. Biol.* **234**, 779–815 (1993).

45. Jo, S., Kim, T., Iyer, V. G. & Im, W. CHARMM-GUI: a web-based graphical user interface for CHARMM. *J. Comput. Chem.* **29**, 1859–1865 (2008).

46. Cheng, X., Jo, S., Lee, H. S., Klauda, J. B. & Im, W. CHARMM-GUI Micelle Builder for Pure/Mixed Micelle and Protein/Micelle Complex Systems. *J. Chem. Inf. Model.* **53**, 2171–2180 (2013).

47. Lee, J. *et al.* CHARMM-GUI Input Generator for NAMD, GROMACS, AMBER, OpenMM, and CHARMM/OpenMM Simulations Using the CHARMM36 Additive Force Field. *J. Chem. Theory Comput.* **12**, 405–413 (2016).

48. Lomize, M. A., Pogozheva, I. D., Joo, H., Mosberg, H. I. & Lomize, A. L. OPM database and PPM web server: resources for positioning of proteins in membranes. *Nucleic Acids Res.* **40**, D370-376 (2012).

49. Vanommeslaeghe, K. *et al.* CHARMM general force field: A force field for drug-like molecules compatible with the CHARMM all-atom additive biological force fields. *J. Comput. Chem.* **31**, 671–690 (2010).

50. Kim, S. *et al.* CHARMM-GUI ligand reader and modeler for CHARMM force field generation of small molecules. *J. Comput. Chem.* **38**, 1879–1886 (2017).

51. Guerra, F., Adam, S. & Bondar, A.-N. Revised force-field parameters for chlorophyll-*a*, pheophytin-*a* and plastoquinone-9. *J. Mol. Graph. Model.* **58**, 30–39 (2015).

52. da Silva, T. U., Pougy, K. de C., Albuquerque, M. G., da Silva Lima, C. H. & Machado, S. de P. Development of parameters compatible with the CHARMM36 force field for [Fe4S4]2+ clusters and molecular dynamics simulations of adenosine-5′-phosphosulfate reductase in GROMACS 2019. *J. Biomol. Struct. Dyn.* **40**, 3481–3491 (2022).

53. Larsson, P., Kneiszl, R. C. & Marklund, E. G. MkVsites: A tool for creating GROMACS virtual sites parameters to increase performance in all-atom molecular dynamics simulations. *J. Comput. Chem.* **41**, 1564–1569 (2020).

54. Abraham, M. J. *et al.* GROMACS: High performance molecular simulations through multi-level parallelism from laptops to supercomputers. *SoftwareX* **1–2**, 19–25 (2015).

55. Kohnke, B., Kutzner, C. & Grubmüller, H. A GPU-Accelerated Fast Multipole Method for GROMACS: Performance and Accuracy. *J. Chem. Theory Comput.* **16**, 6938–6949 (2020).

56. Jorgensen, W. L., Chandrasekhar, J., Madura, J. D., Impey, R. W. & Klein, M. L. Comparison of simple potential functions for simulating liquid water. *J. Chem. Phys.* **79**, 926–935 (1983).

57. Bussi, G., Donadio, D. & Parrinello, M. Canonical sampling through velocity rescaling. *J. Chem. Phys.* **126**, 014101 (2007).

58. Essmann, U. *et al.* A smooth particle mesh Ewald method. *J. Chem. Phys.* **103**, 8577–8593 (1995).

59. McGibbon, R. T. *et al.* MDTraj: A Modern Open Library for the Analysis of Molecular Dynamics Trajectories. *Biophys. J.* **109**, 1528–1532 (2015).

60. Persson, L. J., Sahin, C., Landreh, M. & Marklund, E. G. High-Performance Molecular Dynamics Simulations for Native Mass Spectrometry of Large Protein Complexes with the Fast Multipole Method. *Anal. Chem.* **96**, 15023–15030 (2024).

61. Maia, F. R. The coherent X-ray imaging data bank. *Nat. Methods* **9**, 854–855 (2012).

## Acknowledgements

We acknowledge the European XFEL in Schenefeld, Germany, for the provision of X-ray free-electron laser beam time at the SQS instrument (proposal 4098) and thank the European XFEL staff for their assistance. We acknowledge financial support by Deutsches Elektronen-Synchrotron DESY, a member of the Helmholtz Association (HGF). This work was supported by the European Research Council (ERC Consolidator Grant 101088426 to F.R.N.C.M., and ERC Synergy Grant AXSIS 609920 to H.N.C.); the Swedish Research Council (grant nos. 2018-00234 to F.R.N.C.M., 2020-04825 to E.G.M. and L.J.P., 2019-03935 to N.T., 2018-00740 and 2021-05988 to C.C., and 2022-00236 to A.M.); the Röntgen-Ångström Cluster (grant nos. 2019-06092 and 2023-06348 to F.R.N.C.M., and 2023-06350 to J.A.S.); the Czech Science Foundation's GACR (project no. 24-10671S to J.H.); the Johannes Amos Comenius Operational Programme (OPJAK) financed by the European Structural and Investment Funds and the Czech Ministry of Education, Youth and Sports for the SENDISO project (no. CZ.02.01.01/00/22_008/0004596 to J.H.); the Human Frontier Science Program (grant no. RGP0010/2017 to P.L.X. and H.N.C.); the Marsden Fund Council, managed by Royal Society Te Apārangi (grant no. MFP-UOC2203 to M.D. and J.P.J.C.); the Deutsche Forschungsgemeinschaft (DFG) under Germany's Excellence Strategy – EXC 2056 – Project ID 390715994 (Cluster of Excellence 'CUI: Advanced Imaging of Matter' supporting K.A., J.K., and T.W.); and the Norwegian Research Council (grant no. NFR_335017 to L.A.E.). P.L.X. thanks the Joachim Herz Stiftung for a fellowship.

## Author contributions

F.R.N.C.M. conceived and designed the experiment. F.R.N.C.M., K.A., and T.V.Y. wrote the manuscript, with contributions from E.G.M., L.J.P., N.T., C.C., L.A.E., P.L.X., P.A. and J.H.. All authors discussed the results and provided feedback on the manuscript.

E.G.M., L.J.P., J.B., A.K.S., L.W., J.K., P.K., H.N.C., J.H., and T.V.Y. contributed to experimental design. E.G.M., L.J.P., N.T., C.C., A.B., and S.C. performed numerical modelling. N.T., M.D., J.P.J.C., J.A.S., K.A., A.B., A.J.M., and S.C. performed online data analysis. L.J.P., K.A., Ab.M., P.M., P.A., Z.S., T.W., L.A.E., A.J.M., and A.W. performed offline data analysis. T.V.Y., M.D., J.P.J.C., J.B., J.A.S., R.A., A.A., D.M., R.A.K., A.K.S., L.W., A.D.E., L.V.H., J.K., P.K., A.S., S.R.-Z., and D.W. contributed to sample delivery. A.M., L.A.E., J.D.M.-A., P.F., P.L.X., and Ch.S. performed sample preparation and characterization. I.d.D.M. and C.S. performed EM sample characterization. M.M., S.U., S.D., R.B., Y.O., A.D.F., and B.S. were responsible for the design and operation of the SQS instrument. D.L., C.D., M.P., and A.C. operated and calibrated the DSSC detector.

## Competing interests

The authors declare no competing interests.

# Three-dimensional imaging of isolated membrane-protein complexes *in vacuo* with an X-ray laser

Kartik Ayyer *et al.*

## Supplementary Information

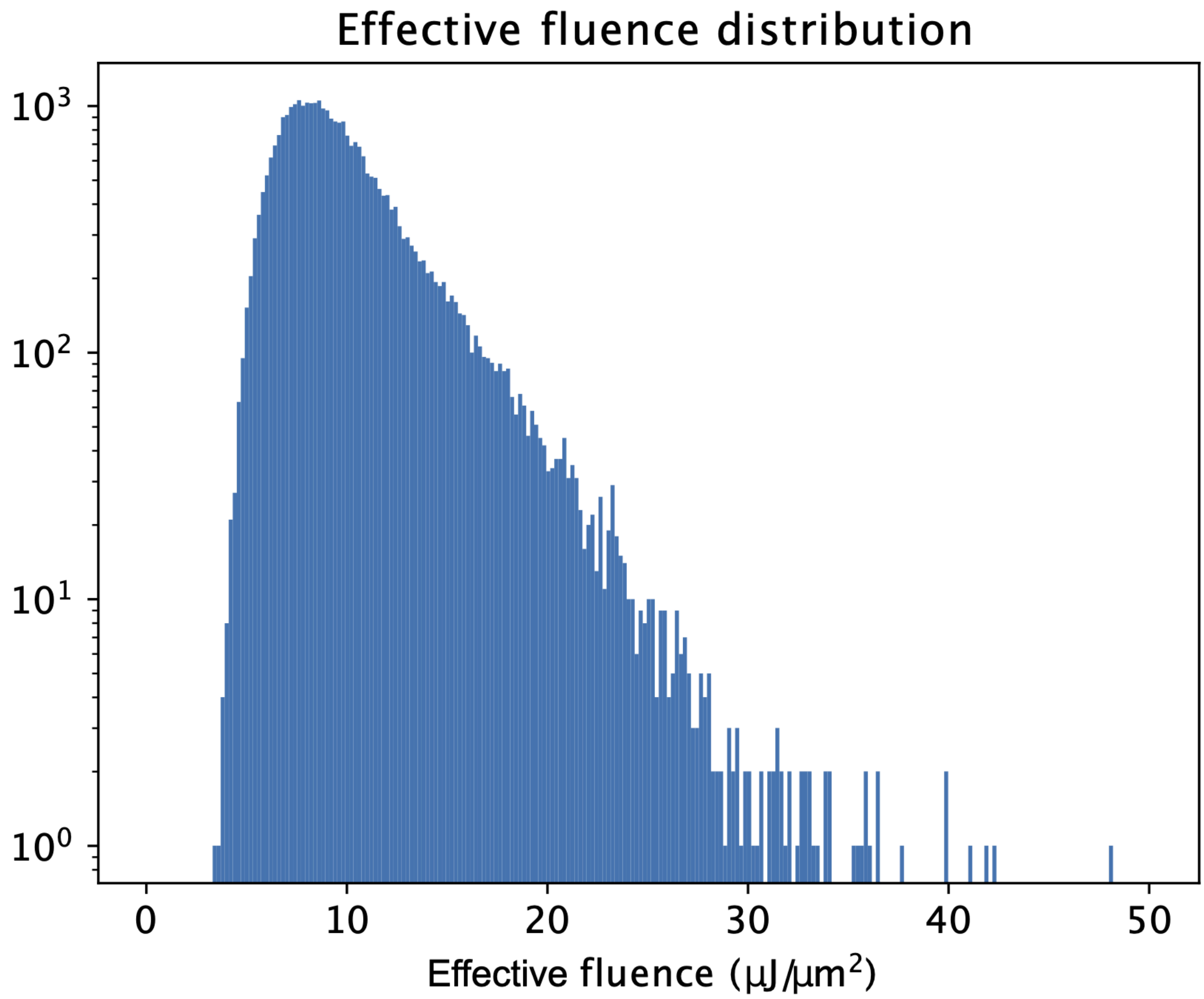

**Fig. S1 | Effective fluence fitted from the diffraction patterns used for Fourier assembly.** Histogram of the number of particles per effective fluence, the fluence that would be required on the sample to produce the observed diffraction patterns if the sample were immune to radiation damage effects.

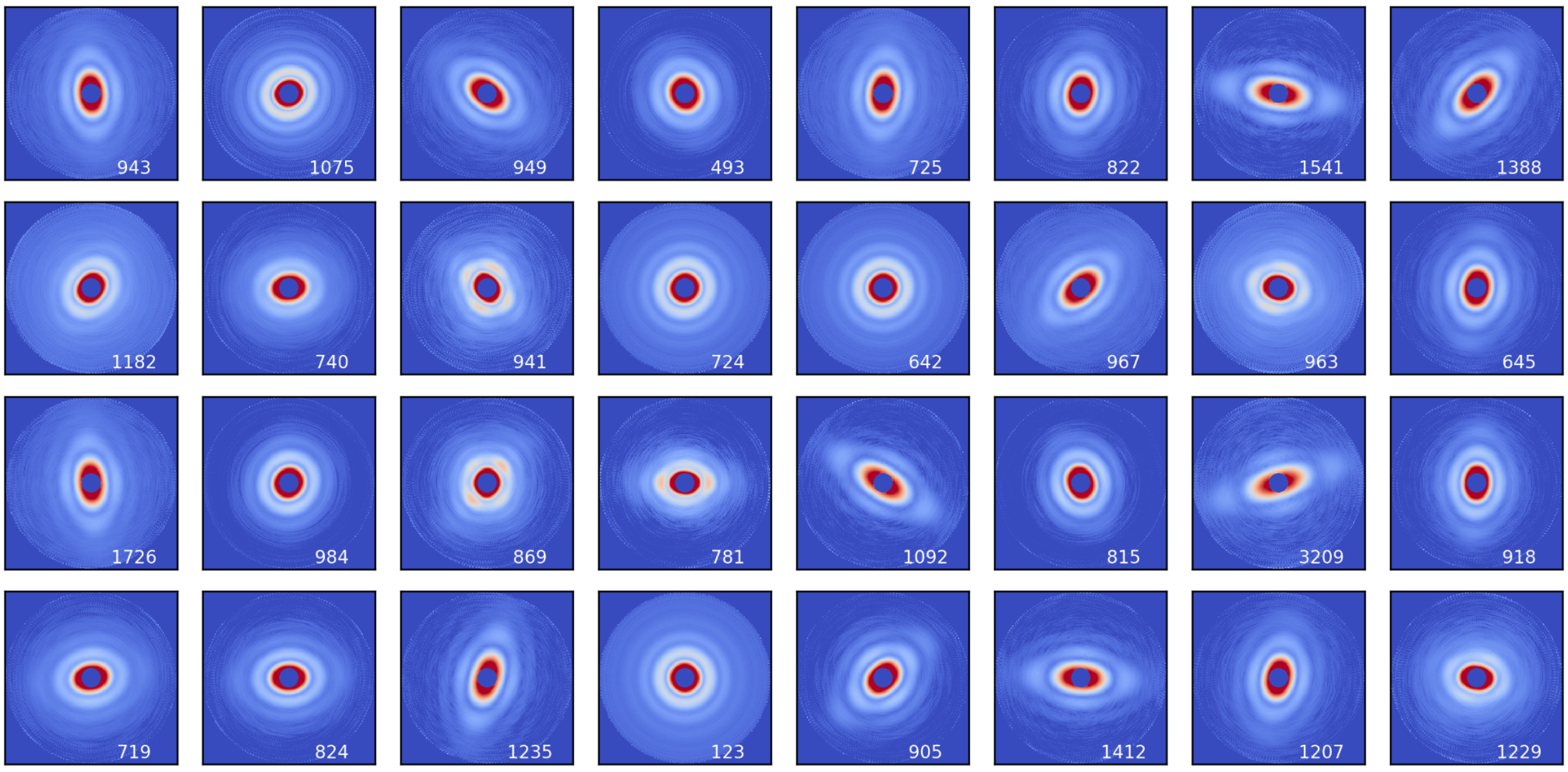

**Fig. S2 | 2D classification of the data that went into the 3D reconstruction.** Two-dimensional class averages of the 32 788 patterns used in the 3D reconstruction. The number in the corner of each class corresponds to the number of patterns in it.

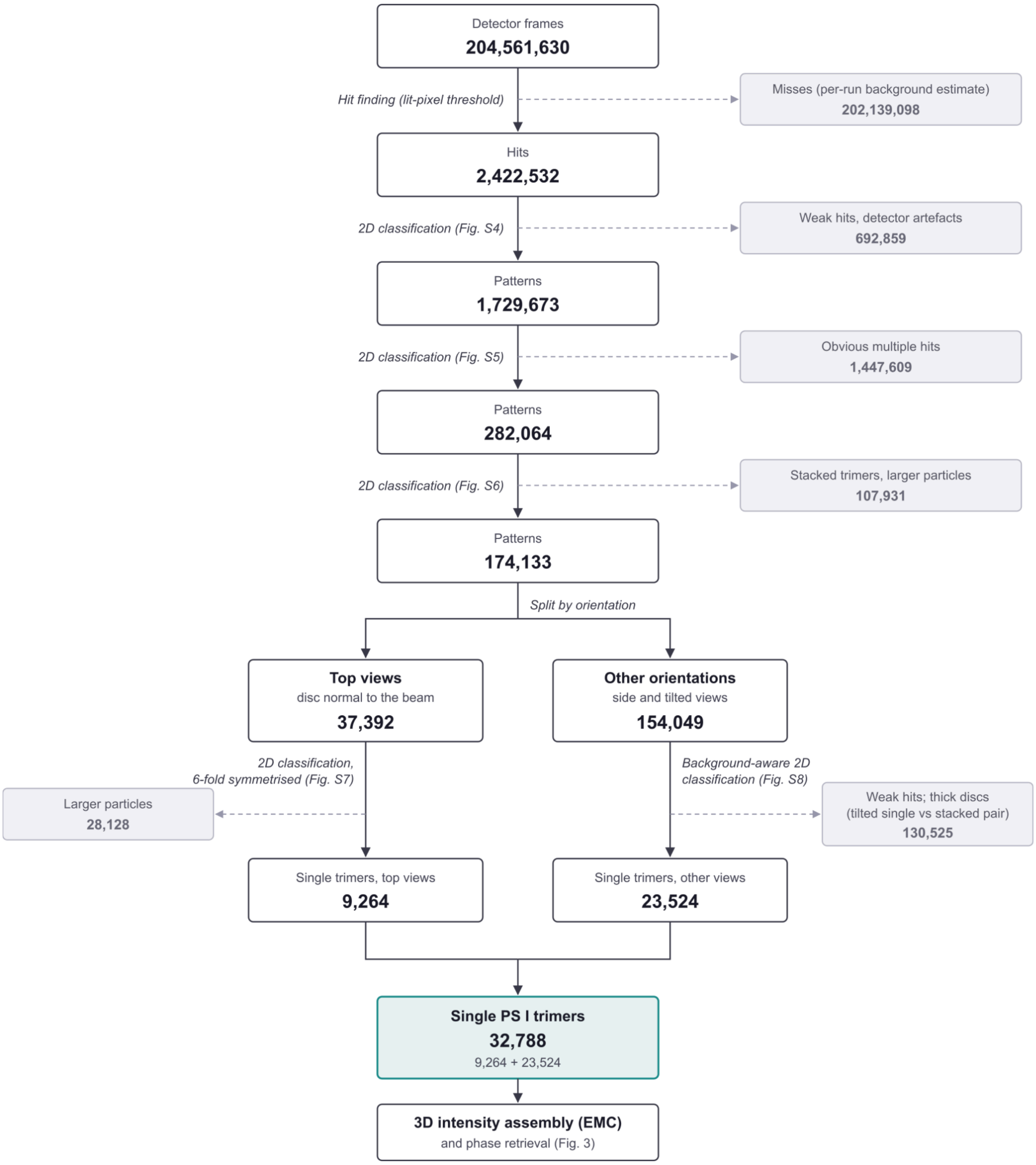

**Fig. S3 | Flow diagram of the data and filtering used.** Flow of detector frames through hit finding and 2D classification. The main column shows the number of patterns retained after each step; grey boxes on the right show the number rejected and why. Frames without a hit ("misses") were not discarded but used to estimate the background for each run. Hits were classified in three successive rounds to remove weak hits and detector artefacts (Fig. S4), obvious multiple hits (Fig. S5), and stacked trimers and other particles larger than a single trimer (Fig. S6). The remaining patterns were split by orientation and classified separately: top views with six-fold symmetry imposed, keeping only the smallest particles (Fig. S7), and all other orientations with background-aware classification, rejecting weak hits and thick discs that could be either tilted single trimers or edge-on stacked pairs (Fig. S8). The two

selections were combined into the 32 788 patterns used for 3D intensity assembly and phase retrieval (Fig. 3).

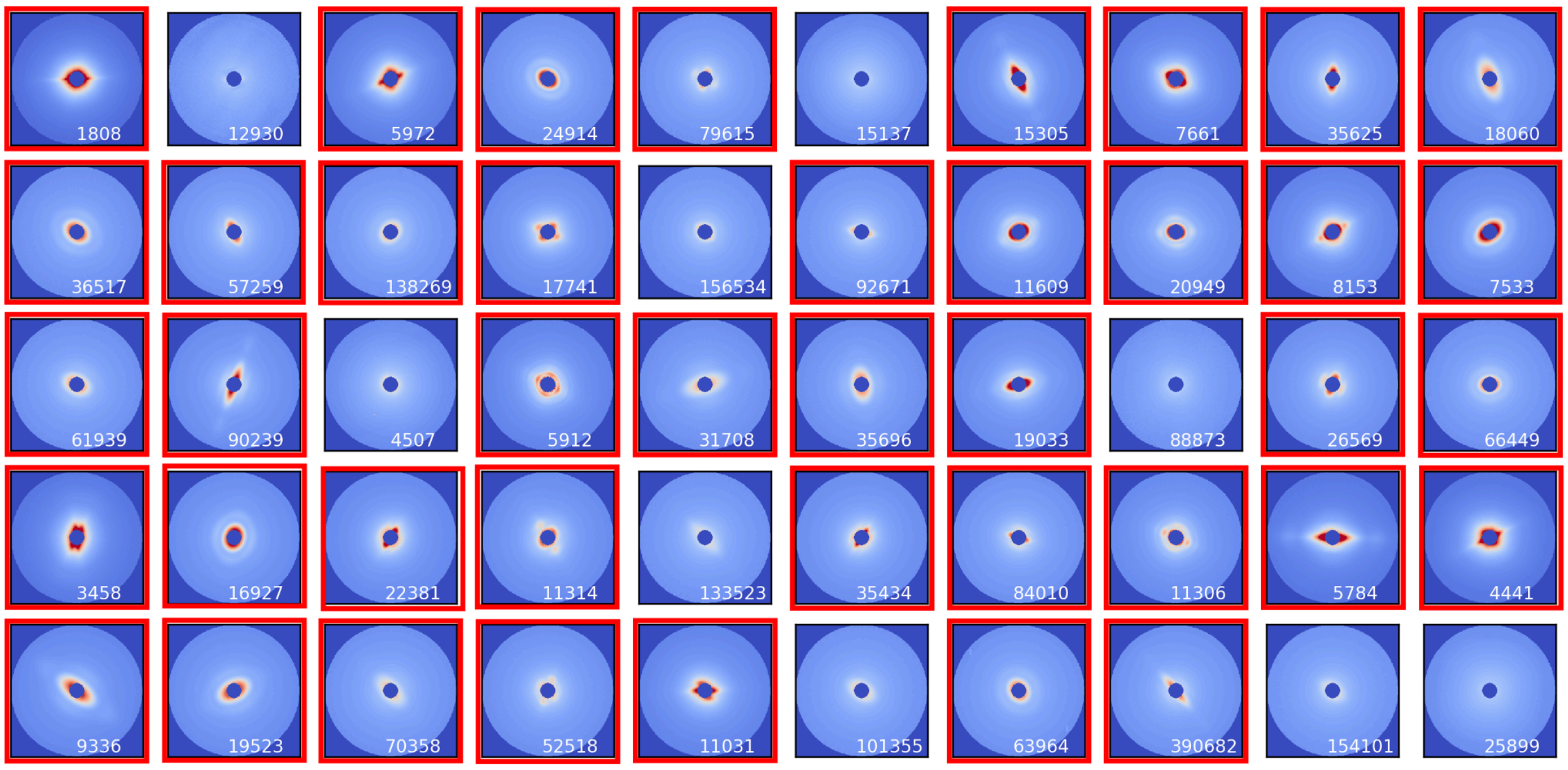

**Fig. S4 | First round of 2D classification.** Two-dimensional class averages of the 2 422 532 patterns used in the first round of 2D classification. Classes dominated by weak hits or detector artefacts were discarded resulting in 1 729 673 patterns. The number in the corner of each class corresponds to the number of patterns in it and the red squares mark the classes retained for subsequent rounds of classification.

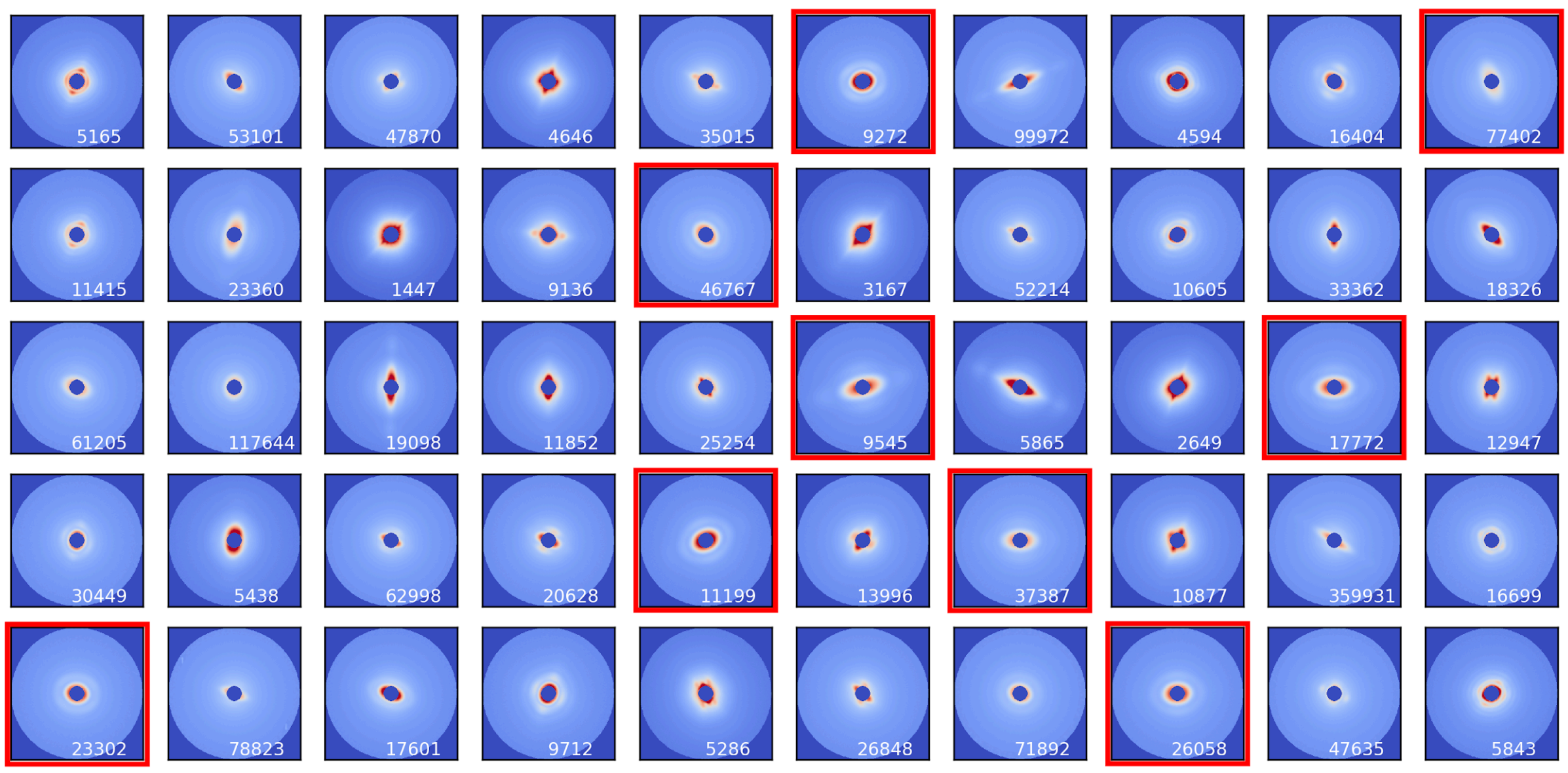

**Fig. S5 | Second round of 2D classification.** Two-dimensional class averages of the 1 729 673 patterns used in the second round of 2D classification. Classes showing obvious multiple hits, showing high frequency modulation, were removed giving a total of 282 064 images. As

before, the number in the corner of each class corresponds to the number of patterns in it and the red squares mark the classes retained for subsequent rounds of classification.

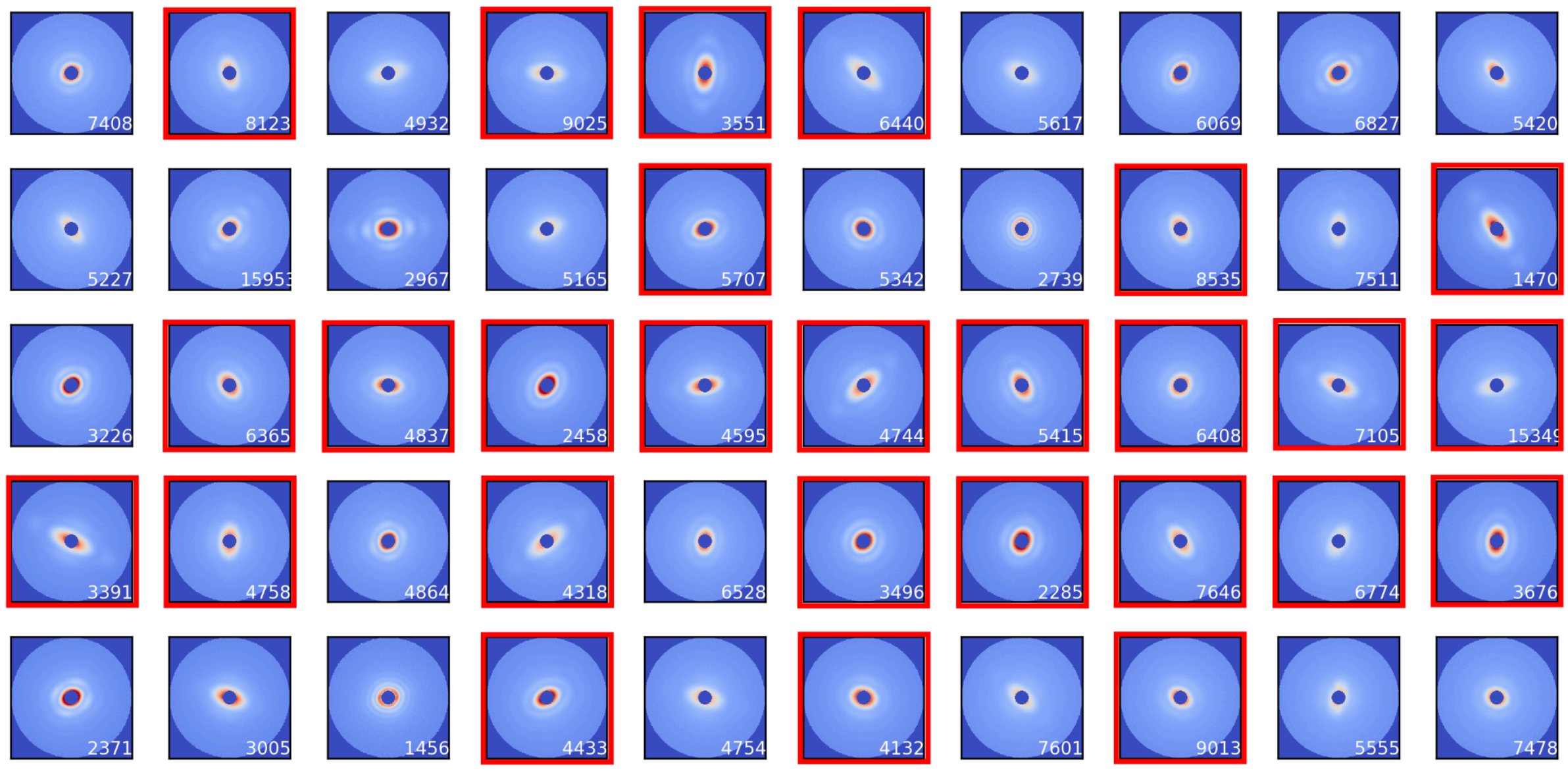

**Fig. S6 | Third round of 2D classification.** Two-dimensional class averages of the 282 064 selected the second round of 2D classification. Classes consistent with stacked trimers and other particles larger than a single micelle-embedded trimer were removed leaving 174 133 patterns.

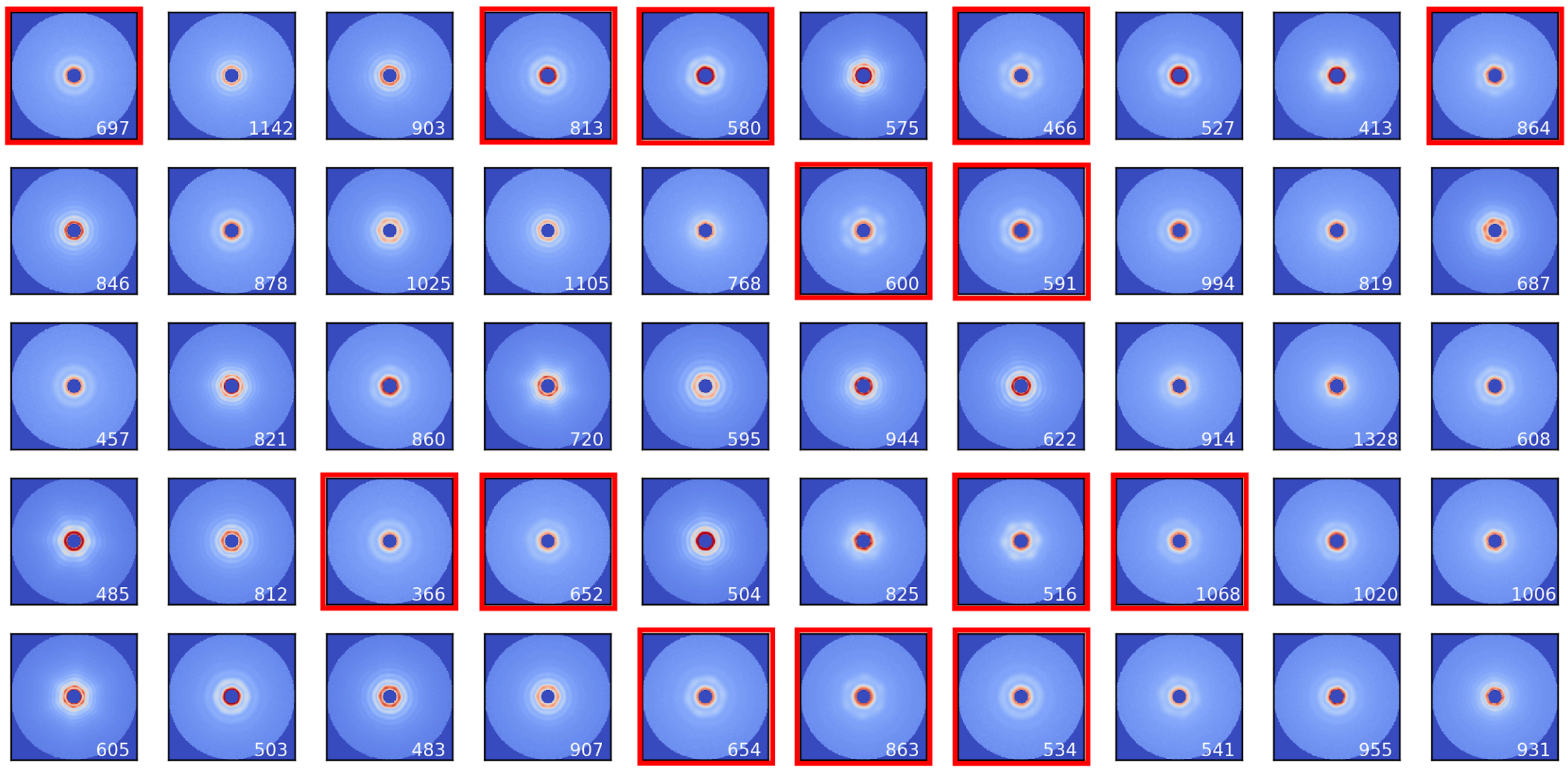

**Fig. S7 | 2D classification of top views.** Two-dimensional class averages of the 37 392 images selected in the third round of 2D classification that correspond to top views. These patterns were classified separately with six-fold symmetry imposed, as expected from the trimer's three-fold axis combined with the centrosymmetry of the diffraction intensities. Only classes with the largest fringe spacing (the smallest particles) were kept, giving 9 264 patterns.

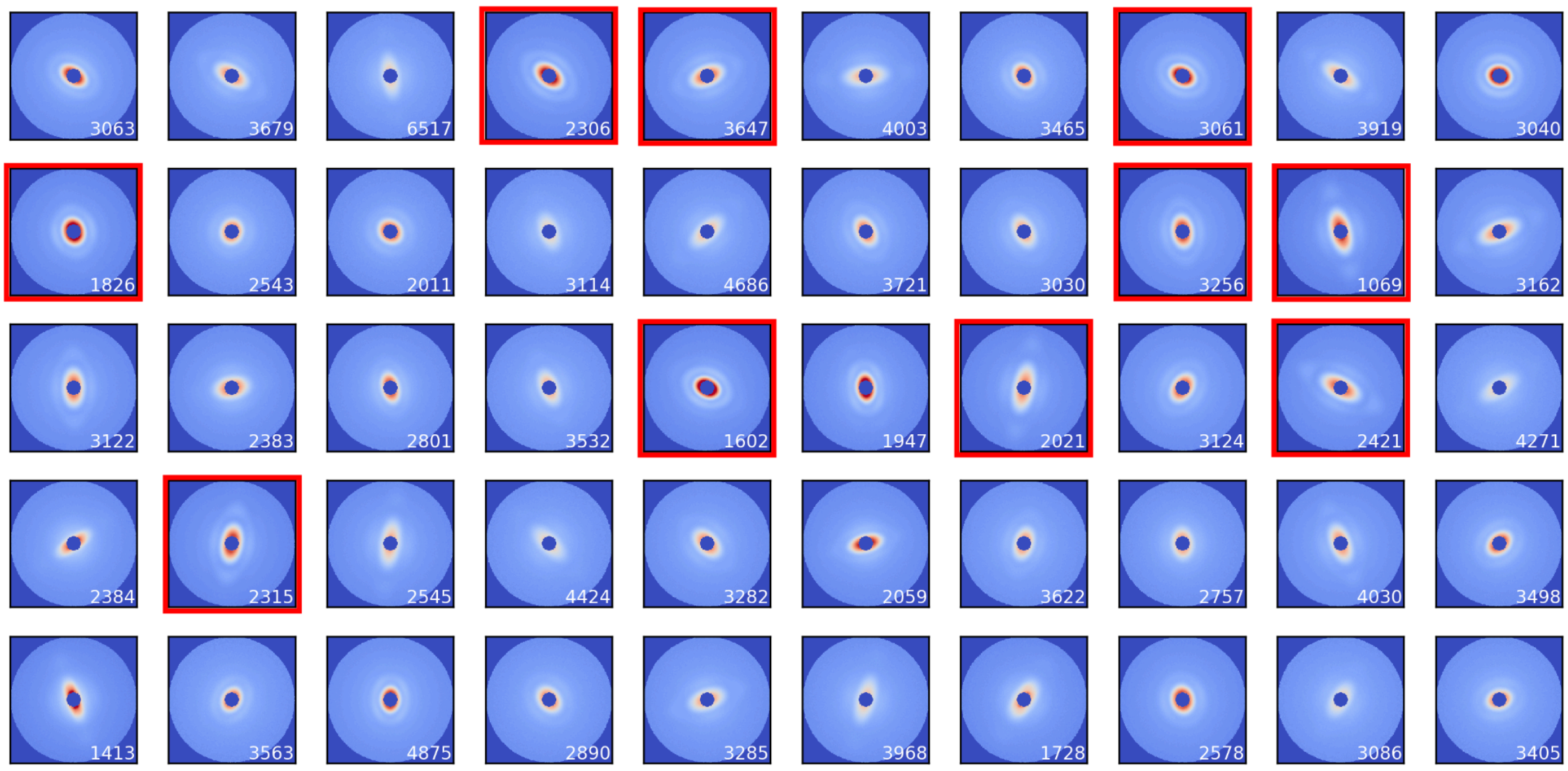

**Fig. S8 | 2D classification of other views.** Two-dimensional class averages of the 154 049 selected the third round of 2D classification that are not top views. Classes of weak hits were rejected, as were classes corresponding to thicker discs. A tilted single trimer and an edge-on pair of stacked trimers give similar patterns, so these ambiguous classes were excluded conservatively, resulting in 23 524 patterns.

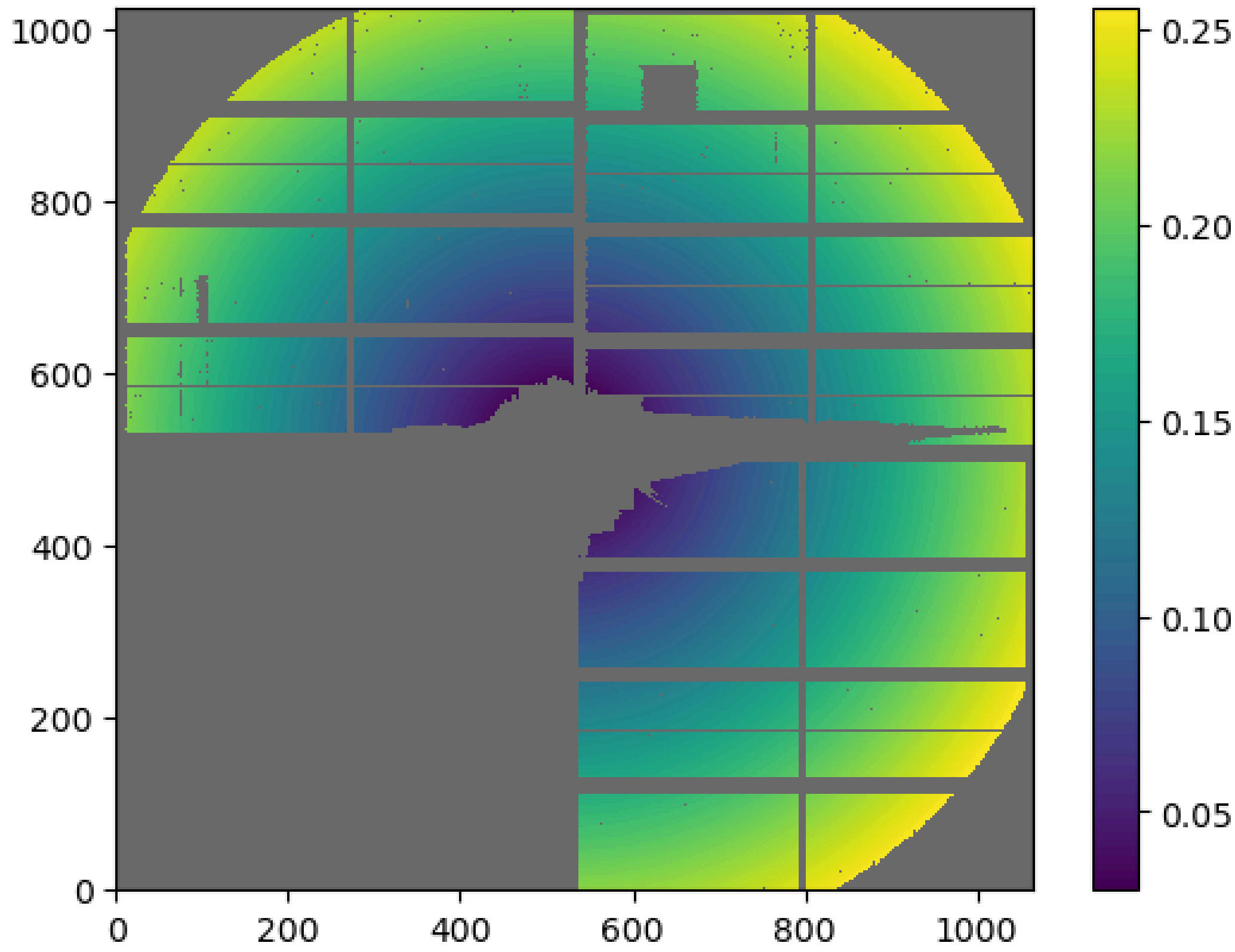

**Fig. S9 | Usable detector area.** Pixels used in the analysis are coloured by momentum transfer *1/d* ($nm^{-1}$). Grey pixels are masked. The usable area extends from $q \approx 0.03\ nm^{-1}$ to $0.26\ nm^{-1}$, corresponding to resolutions of about 33 nm to 3.8 nm, at a detector distance of 536 mm and a photon energy of 1.2 keV. The outer circular boundary is set by the shadow of the flange holding the detector. Also masked are the lower-left quadrant, which was not operational during the experiment; the region around the direct beam, including the streak extending to the right of centre, due to upstream scattering; the gaps between sensor modules and individual defective pixels and rows. Axes give pixel indices.